\documentclass[11pt,a4paper]{article}
\pdfoutput=1
\usepackage{jheppub}

\usepackage[numbers,sort&compress]{natbib}
\usepackage{subcaption}
\usepackage{enumerate}
\usepackage{xifthen} 
\usepackage{tikz}

\renewcommand*{\le}{\left}
\newcommand*{\ri}{\right}

\renewcommand*{\a}{\alpha}
\renewcommand*{\b}{\beta}
\newcommand*{\g}{\gamma}
\renewcommand*{\d}{\delta}
\newcommand*{\e}{\epsilon}

\newcommand*{\q}{\theta}
\renewcommand*{\l}{\lambda}

\renewcommand*{\r}{\rho}

\renewcommand*{\t}{\tau}
\newcommand*{\f}{\phi}
\newcommand*{\vf}{\varphi}
\renewcommand*{\k}{\kappa}

\newcommand*{\y}{\psi}
\newcommand*{\w}{\omega}
\newcommand*{\wb}{\bar{\omega}}
\newcommand*{\F}{\Phi}

\newcommand*{\p}{\partial}
\renewcommand{\Re}{\operatorname{Re}} 
\renewcommand{\Im}{\operatorname{Im}} 

\newcommand*{\cF}{\mathcal{F}}

\newcommand*{\cN}{\mathcal{N}}
\newcommand*{\cO}{\mathcal{O}}
\newcommand*{\cZ}{\mathcal{Z}}

\newcommand*{\C}[2][]{
\ifthenelse{\isempty{#1}}{
{C_{#2}}
}{
	{C_{#2,\,#1}}
}
}
\newcommand*{\CF}[2][]{
\ifthenelse{\isempty{#1}}{
{F_{#2}}
}{
	{F_{#2,\,#1}}
}
}

\title{Holographic Coulomb Branch Solitons, Quasinormal Modes, and Black Holes}
\author[1]{S. Prem Kumar,}
\author[2]{Andy O'Bannon,}
\author[3]{Anton Pribytok,}
\author[4]{Ronnie Rodgers,}
\affiliation[1]{Department of Physics, Swansea University, Swansea, SA2 8PP, United Kingdom}
\affiliation[2]{STAG Research Centre, Physics and Astronomy, University of Southampton,\\
    Highfield, Southampton SO17 1BJ, United Kingdom}
\affiliation[3]{School of Mathematics \& Hamilton Mathematics Institute, Trinity College Dublin, Dublin, Ireland}
\affiliation[4]{Institute  for  Theoretical  Physics,  Utrecht  University, 3584  CC  Utrecht,  The  Netherlands}
\emailAdd{s.p.kumar@swansea.ac.uk}
\emailAdd{a.obannon@soton.ac.uk}
\emailAdd{apribytok@maths.tcd.ie}
\emailAdd{r.j.rodgers@uu.nl}
\abstract{
	Four-dimensional $\mathcal{N}=4$ supersymmetric Yang-Mills theory, at a point on the Coulomb branch where $SU(N)$ gauge symmetry is spontaneously broken to $SU(N-1)\times U(1)$, admits BPS solitons describing a spherical shell of electric and/or magnetic charges enclosing a region of unbroken gauge symmetry. These solitons have been proposed as gauge theory models for certain features of asymptotically flat extremal black holes. In the 't Hooft large $N$ limit with large 't Hooft coupling, these solitons are holographically dual to certain probe D3-branes in the $AdS_5 \times S^5$ solution of type IIB supergravity. By studying linearised perturbations of these D3-branes, we show that the solitons support quasinormal modes with a spectrum of frequencies sharing both qualitative and quantitative features with asymptotically flat extremal black holes.
}

\begin{document}

\maketitle

\section{Introduction}
\label{sec:intro}

In the effort to understand quantum gravity, black holes remain a mystery of central importance. What microstates contribute to their Bekenstein-Hawking entropy? Why does that entropy scale with area, not volume? How does quantum gravity resolve the information loss paradox? Can the information apparently lost when degrees of freedom fall into a black hole somehow be recovered by observers outside the black hole?

One possible route to a better understanding of black holes is to study objects with similar properties in quantum field theory (QFT). Schwarz proposed one such object in refs.~\cite{Schwarz:2014rxa,Schwarz:2014zsa}: dyonic BPS solitons of $(3+1)$-dimensional \(\cN=4\) supersymmetric Yang-Mills theory (SYM) with gauge group $SU(N)$, at a point on the Coloumb branch where a non-zero vacuum expectation value (VEV) of one adjoint-valued scalar field spontaneously breaks the gauge group down to \(SU(N-1) \times U(1)\).

\begin{figure}
   \begin{subfigure}[t]{0.5\textwidth}
   \begin{center}
        \begin{tikzpicture}
           \draw[thick] (0,-1.5)--(1,-0.5)--(5,-0.5)--(4,-1.5)--(0,-1.5);
           \draw[thick] (0,-3.2)--(1,-2.2)--(5,-2.2)--(4,-3.2)--cycle;
           \draw[thick,fill=white] (0,-3.1)--(1,-2.1)--(5,-2.1)--(4,-3.1)--cycle;
           \draw[thick,fill=white] (0,-3)--(1,-2)--(5,-2)--(4,-3)--cycle;
           \draw[thick, <->] (-0.1,-3)--(-0.1,-1.5) node[midway,right]{\(\Delta\)};
           \node at (5.7,-2.1) {\(\}N-1\)};
        \end{tikzpicture}
        \caption{Coulomb branch in flat space}
        \label{fig:cartoon_coulomb_branch}
   \end{center}
   \end{subfigure}
   \begin{subfigure}[t]{0.5\textwidth}
   \begin{center}
        \begin{tikzpicture}
            \draw[thick,->] (0,-2.5) -- (0,0) node[left]{\(r\)};
            \node[left] at (0,-1) {\(\Delta\)};
            \node[left] at (0,-2.5) {\(0\)};
            \draw[thick,dotted] (0,-2.5) -- (5,-2.5);
            \draw[thick] (0,-1) -- (5,-1);
            \draw[thick] (2.5,-2.6) -- (2.5,-2.4);
            \draw[thick,->] (2.5,-2.5) -- (3.2,-2.5) node[below]{\(\r\)};
            \node[below] at (2.5,-2.5) {\(0\)};
        \end{tikzpicture}
        \caption{Coulomb branch in \(AdS_5\)}
        \label{fig:coulomb_branch_ads}
   \end{center}
   \end{subfigure}
   \begin{subfigure}[t]{0.5\textwidth}
   \begin{center}
        \begin{tikzpicture}
           \draw[thick] (1,-1) to[out=0,in=90] (2.3,-2.5);
           \draw[thick] (2.7,-2.5) to[out=90,in=180] (4,-1);
           \draw[thick] (0,-3.2)--(1,-2.2)--(5,-2.2)--(4,-3.2)--cycle;
           \draw[thick,fill=white] (0,-3.1)--(1,-2.1)--(5,-2.1)--(4,-3.1)--cycle;
           \draw[thick,fill=white] (0,-3)--(1,-2)--(5,-2)--(4,-3)--cycle;
           \draw[thick, <->] (-0.1,-3)--(-0.1,-1.5) node[midway,right]{\(\Delta\)};
           \node at (5.7,-2.1) {\(\}N-1\)};
           \fill[white] (1,-1) to[out=0,in=90] (2.3,-2.5) -- (2.7,-2.5) to[out=90,in=180] (4,-1) -- cycle;
           \draw[thick] (1,-1) to[out=0,in=90] (2.3,-2.5) to[out=-90,in=0] (2.2,-2.6);
           \draw[thick] (2.8,-2.6) to[out=180,in=-90] (2.7,-2.5) to[out=90,in=180] (4,-1);
           \fill[white] (1.75,-1.5) -- (1.73,-1.25) to[out=-20,in=200] (3.27,-1.25) -- (3.25,-1.5) -- cycle;
           \draw[thick] (0,-1.5)--(1,-0.5)--(5,-0.5)--(4,-1.5)--(0,-1.5);
           \draw[thick] (1.72,-1.25) to[out=-20,in=200] (3.28,-1.25);
           \draw[thick,dotted] (1.725,-1.25) to[out=20,in=150] (3.275,-1.25);
        \end{tikzpicture}
        \caption{Dyonic soliton in flat space}
        \label{fig:cartoon_soliton}
   \end{center}
   \end{subfigure}
   \begin{subfigure}[t]{0.5\textwidth}
   \begin{center}
        \begin{tikzpicture}
            \draw[thick,->] (0,-2.5) -- (0,0) node[left]{\(r\)};
            \node[left] at (0,-1) {\(\Delta\)};
            \node[left] at (0,-2.5) {\(0\)};
            \draw[thick,dotted] (0,-2.5) -- (5,-2.5);
            \draw[thick] (0,-1) to[out=0,in=110] (2.3,-2.5);
            \draw[thick] (5,-1) to[out=180,in=70] (2.7,-2.5);
            \draw[thick] (2.5,-2.6) -- (2.5,-2.4);
            \draw[thick,->] (2.5,-2.5) -- (3.2,-2.5) node[below]{\(\r\)};
            \node[below] at (2.5,-2.5) {\(0\)};
        \end{tikzpicture}
        \caption{Dyonic soliton in \(AdS_5\)}
        \label{fig:soliton_ads}
   \end{center}
   \end{subfigure}
   \caption{Cartoons of the D3-brane configurations discussed in this paper.}
   \label{fig:cartoons}
\end{figure}
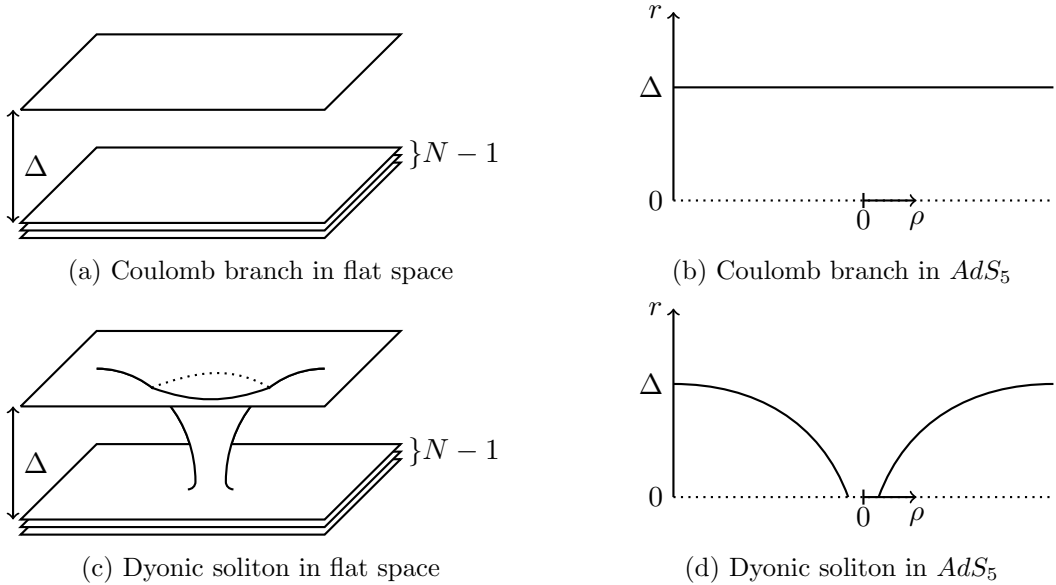

The solitons in question may be understood from the following string theory construction. Consider a flat stack of \(N\) coincident D3-branes in $(9+1)$-dimensional Minkowski space. The low-energy world volume theory on these D3-branes is \(\cN=4\) SYM with gauge group \(SU(N)\), where the fields of \(\cN=4\) SYM arise from open strings beginning and ending on the stack of D3-branes. If one brane is separated from the stack by a distance \(\Delta\), as depicted in figure~\ref{fig:cartoon_coulomb_branch}, then one of the scalar fields of \(\cN=4\) SYM develops a VEV proportional to \(\Delta\), spontaneously breaking the gauge group to \(SU(N-1) \times U(1)\). Since this is a point on the Coulomb branch of \(\cN=4\) SYM, we will refer to the separated D3-brane as the Coulomb branch D3-brane.

In the 't Hooft large-\(N\) limit with large 't Hooft coupling \(\l\gg1\), \(\cN=4\) SYM with gauge group \(SU(N)\) is holographically dual to type IIB supergravity on \(AdS_5 \times S^5\), with $N$ units of five-form flux on the $S^5$~\cite{Maldacena:1997re,Witten:1998qj,Gubser:1998bc}, which is the near-horizon limit of the geometry sourced by the stack of D3-branes. The spontaneous breaking \(SU(N) \to SU(N-1) \times U(1)\) is described by the embedding of a probe D3-brane into this geometry. Concretely, we will work in coordinates in which the \(AdS_5 \times S^5\) metric is
\begin{equation}
\label{eq:background_metric}
   d s^2 =  \frac{L^2}{r^2} dr^2+\frac{r^2}{L^2} \le(
     - d t^2 + d \r^2  + \r^2 d \q^2 + \r^2 \sin^2 \q \, d\vf^2
  \ri)
    + L^2 d s_{S^5}^2,
\end{equation}
where $L$ is the $AdS_5$ radius of curvature, $r$ is the $AdS_5$ radial coordinate, with Poincar\'e horizon at $r\to0$ and boundary at $r \to \infty$, $(t,\rho,\theta,\varphi)$ are the time and spherical coordinates of $\cN=4$ SYM, and \(ds_{S^5}^2\) is the metric of a unit round \(S^5\). The Coulomb branch D3-brane spans \((t,\r,\q,\vf)\) and sits at \(r=\Delta\) and at an arbitrary point on the \(S^5\). The projection of this embedding onto the \((\r,r)\) plane is sketched in fig.~\ref{fig:coulomb_branch_ads}. Schwarz has conjectured that the action of the probe D3-brane in \(AdS_5 \times S^5\) provides the effective action for the massless \(U(1)\) sector fields. In fact, Schwarz claimed this was a ``highly effective action'' (HEA)~\cite{Schwarz:2013wra}: unlike a usual low-energy effective action, the HEA would be valid at all energy scales.

For the D3-branes in flat space the W-bosons, which gain mass $M_W$ from the gauge symmetry breaking, correspond to open strings stretched between the Coulomb branch brane and the stack. Similarly, a magnetic monopole corresponds to a stretched D1-brane. In \(AdS_5 \times S^5\), the W-bosons and magnetic monopoles correspond to strings and D1-branes that stretch from the probe D3-brane to the Poincar\'e horizon at \(r=0\).

A dyonic soliton corresponds to a \((p,q)\)-string --- a bound state of \(p\) strings and \(q\) D1-branes --- stretched between the branes. When at least one of \(p\) or \(q\) is sufficiently large, the \((p,q)\)-string deforms the D3-branes~\cite{Callan:1997kz,Gibbons:1997xz}, producing a throat that connects the Coulomb branch brane to the stack, as sketched in fig.~\ref{fig:cartoon_soliton}. In \(AdS_5 \times S^5\), in the coordinates of eq.~\eqref{eq:background_metric}, the soliton corresponds to a D3-brane embedding that again spans \((t,\r,\q,\vf)\),  but now with \(r\) that depends on \(\r\) as~\cite{Schwarz:2014rxa,Schwarz:2014zsa}\footnote{Similar solutions in the full asymptotically flat D3-brane geometry, of which \(AdS_5 \times S^5\) is the near-horizon limit, were found in ref.~\cite{Gauntlett:1999xz}.}
\begin{equation}
\label{eq:r_rho}
   r(\r) = \Delta - \frac{L^2 \k}{\r},
\end{equation}
where \(\k\) is a dimensionless constant fully determined by \(p\) and \(q\). The D3-brane also carries electric and magnetic fields pointing in the \(\r\) direction, sourced by the \(p\) strings and \(q\) D1-branes, respectively. We review the construction of this solution in section~\ref{sec:solution}, where we also give the precise relation between \(\k\), \(p\), and \(q\). The qualitative form of \(r(\r)\) is sketched in fig.~\ref{fig:soliton_ads}. As \(\r\to\infty\), the brane asymptotes to the Coulomb branch embedding \(r = \Delta\). As \(\rho\) decreases from infinity, \(r(\r)\) decreases, eventually reaching \(r=0\) at \(\r = \r_0 = L^2 \k/\Delta\). Since \(r\) cannot be negative, the D3-brane is only present for \(\r \geq \r_0\).

What is the interpretation of the soliton solution in the dual QFT? The value of \(r(\r)\) is proportional to the VEV of the scalar field responsible for the gauge symmetry breaking. The VEV thus varies with distance from the soliton, interpolating between a constant proportional to \(\Delta\) at large distances, down to zero at a spherical shell of radius \(\r_0\). Inside the shell, the gauge symmetry is unbroken \(SU(N)\). The radial electric and magnetic fields of the probe D3-brane translate directly into radial electric and magnetic fields of the \(U(1)\) sector in the symmetry-broken phase. The shell at \(\r_0\) therefore carries both electric charge \(p\) and magnetic monopole charge \(q\). We note at this stage that the monopoles described by the D3-brane solutions should be interpreted as non-Abelian GNO~\cite{Goddard:422230} monopoles, since the gauge symmetry is broken to $SU(N-1)\times U(1)$ by the adjoint Higgs mechanism.

Similar BPS soliton solutions, describing an electrically and magnetically charged spherical shell, were found by Popescu and Shapere in \(\cN=2\) pure SYM with gauge group \(SU(2)\) broken to \(U(1)\)~\cite{Popescu:2001rf}. In this case, the scalar component of the \(\cN=2\) vector multiplet in the \(U(1)\) sector varies with distance outside the shell, while taking a constant (but generically non-zero) value inside the shell. These solutions may be the gravitational decoupling limit of the gravitational ``empty holes'' studied in refs.~\cite{Denef:2000nb,Denef:2001xn}. Spherical shells of charge also appear in other D-brane constructions: see for example refs.~\cite{deMelloKoch:1999ui,Johnson:1999qt}.

The $\cN=4$ SYM soliton's total mass $M$ and total charge $Q$ are both proportional to its radius, \(M, Q \propto \r_0\)~\cite{Schwarz:2014zsa,Schwarz:2014rxa}. This behaviour is exotic compared to the \(\r_0^2\) or \(\r_0^3\) scalings expected for a spherical shell or solid ball, respectively. However, such behavior is similar to extremal Reissner-Nordstr\"om black holes, whose mass and charge are both proportional to the radius of the event horizon. This similarity lead Schwarz to propose that the soliton may provide a QFT model of an asymptotically flat, extremal black hole.~\cite{Schwarz:2014rxa,Schwarz:2014zsa}. At least, the soliton may reproduce certain properties of black holes---but clearly not all. In particular, the QFT soliton has no event horizon, and so cannot describe many fundamental black hole phenomena. Nevertheless, the soliton may have non-zero entropy or other properties similar to black holes, and hence may serve as a ``toy model'' for them.

Indeed, in this paper we demonstrate a further black hole-like property of these solitons: they have quasinormal modes (QNMs). For an asymptotically flat black hole, a QNM is a perturbation that obeys outgoing boundary conditions both at spatial infinity and at the horizon. Solutions obeying these boundary conditions typically only exist at certain complex frequencies. A probe brane in \(AdS\) that extends to \(\r \to \infty\), such as the D3-brane embedding in eq.~\eqref{eq:r_rho}, admits outgoing boundary conditions at infinity~\cite{Evans:2019pcs}. In sec.~\ref{sec:fluctuation_equations} we show that outgoing boundary conditions are also possible at \(r=0\), i.e. that waves can travel from the D3-brane into the Poincar\'e horizon. In fact, near $r=0$ the D3-brane's worldvolume geometry is a warped product of $AdS_2$ and $S^2$, somewhat similar to extremal Reissner-Nordstr\"om's near-horizon $AdS_2 \times S^2$.

The QNM solutions are semiclassical excited states of the probe D3-brane, with a decay lifetime inversely proportional to the imaginary part of the QNM frequency. Via the Anti-de Sitter/Conformal Field Theory (AdS/CFT) correspondence~\cite{Maldacena:1997re,Gubser:1998bc,Witten:1998qj}, these states should map to states of the dual QFT. If the HEA conjecture is true, then this map is very direct. In $AdS_5 \times S^5$, the QNM frequencies appear as poles in the Green's functions of the equations of motion of linearised perturbations of D3-brane worldvolume fields. If the D3-brane action is indeed an HEA, then the D3-brane Green's functions are identically equal to two-point functions of the massless \(U(1)\) sector fields.

In sec.~\ref{sec:results} we present our results for the QNM spectrum, computed both numerically and in a WKB limit of large angular momentum. In sec.~\ref{sec:comparisons} we compare this QNM spectrum to those of other systems. In particular, at large angular momentum we find quantitative similarity with asymptotically flat black holes, namely in the complex frequency plane we find QNMs equally-spaced along branches perpendicular to the imaginary axis. We also argue that the late-time decay of perturbations of the soliton will follow power laws more similar to those of extremal Reissner-Nordstr\"om than of non-extremal black holes.

We can also meaningfully compare our QNMs to those of other objects, in gravity and QFT, that describe charged spherical domain walls or phase boundaries, i.e. charged bubbles of some phase.  For such phase bubbles, QNMs arise not because excitations decay by falling through a horizon, but rather because excitations become trapped inside the bubble, and then ``leak out'' over some characteristic timescale. We thus have reason to expect our QNM spectrum to be more similar to a phase bubble's than to a black hole's. Unfortunately, to our knowledge QNMs have not yet been computed for an object that is both charged and a phase bubble. Nevertheless, in sec.~\ref{sec:comparisons} we compare to two objects that are charged \textit{or} a phase bubble, but not both.

First is gravastars, horizonless gravitational objects proposed as alternatives to black holes~\cite{Mazur:2001fv,Visser:2003ge}. A gravastar consists of a bubble of de Sitter spacetime inside an asymptotically flat Schwarzschild spacetime, where the interpolation between the two occurs via a thin shell of matter with appropriate equation of state. QNMs have been computed for uncharged gravastars, and indeed their differences from black hole QNMs may be crucial for distinguishing the two types of objects observationally~\cite{Chirenti:2007mk,Pani:2009ss}. Their QNM spectra are also very different from ours, as we discuss in sec.~\ref{sec:comparisons}. Charged gravastar solutions have been found~\cite{Horvat:2008ch}, but to our knowledge their QNM spectra have not yet been computed.

Second is a magnetic monopole in \(SU(2)\) Yang-Mills theory coupled to an adjoint-valued scalar field that breaks the gauge group to \(U(1)\). This supports QNMs~\cite{Forgacs:2003yh}, but with a spectrum very different from ours, as we discuss in sec.~\ref{sec:comparisons}. Indeed, our soliton is actually more similar to the ``magnetic bag'' conjectured to form from a cluster of many monopoles~\cite{Bolognesi:2005rk}. In a magnetic bag, the monopoles are distributed around a closed wall of thickness \(\sim 1/M_W\). For \(q\) monopoles, the size of the bag is \(\sim q/M_W\), so in the large \(q\) limit the thickness of the wall is negligible. Inside the bag the adjoint-valued scalar field vanishes and \(SU(2)\) is unbroken, while outside the bag the scalar field is non-zero and only a \(U(1)\) is preserved. Like the dyonic soliton in \(\cN=4\) SYM, the mass and charge of a spherical magnetic bag are both proportional to its radius, and they have thus also been compared to black holes~\cite{Bolognesi:2010xt,Manton:2011vm}. However, their QNM spectrum has not been computed.

Nevertheless, our results lend significant evidence to the emerging picture that horizonless solitonic phase bubbles in gravity or QFT can reproduce key features of black holes, including in particular QNMs. Such phase bubbles clearly deserve further research

Indeed, a key question is whether solitonic phase bubbles ever have an entropy proportional to their surface area, similar to a black hole's Bekenstein-Hawking entropy. If so, then what are the microstates, and do they teach us anything about the quantum gravity microstates that contribute to a black hole? For the $\cN=4$ SYM soliton, Schwarz conjectured that the entanglement entropy of a spherical region concentric with the bubble, and of the same radius, might be proportional to the surface area, after suitable regularisation, and in the large-charge limit. In the companion paper ref.~\cite{entanglement_paper}, we holographically compute the contribution of various probe D3-branes to the entanglement entropy of a spherical region in $\cN=4$ SYM. For the D3-brane describing the phase bubble, we find that Schwarz's entanglement entropy scales not with the surface area, $\rho_0^2$, but approximately as $\rho_0^{1.2}$.

This paper is organised as follows. In sec.~\ref{sec:solution} we review the probe D3-brane solution holographically dual to the BPS soliton of $\cN=4$ SYM on the Coulomb branch. In sec.~\ref{sec:fluctuation_equations} we derive the equations of motion for fluctuations of D3-brane worldvolume fields about the soliton solution, and demonstrate that all of them have the same QNM spectrum, called ``isospectrality.'' In sec.~\ref{sec:results} we present our numerical and WKB results for the QNM spectrum. In sec.~\ref{sec:comparisons} we compare our QNM spectrum to those of asymptotically flat extremal Reissner-Nordstr\"om, uncharged gravastars, and the $SU(2)$ Yang-Mills magnetic monopole. We conclude with a summary and suggestions for future research in sec.~\ref{sec:outlook}. We collect various technical results in appendices~\ref{app:leavers} and~\ref{app:wkb}.

\section{Review: Probe D3-branes}
\label{sec:solution}

In this section, we review the construction of the solitonic solutions described in sec.~\ref{sec:intro}. We work in coordinates in which the \(AdS_5 \times S^5\) solution sourced by a stack of \(N\) D3-branes is given by eq.~\eqref{eq:background_metric}. Parameterising the \(S^5\) by angles \(\y^I\) with $I=1,\ldots,5$, such that \(ds_{S^5}^2 = (d\y^1)^2 + \sin^2 \y^1 \, (d\y^2)^2 + \dots\;\), we choose a gauge in which the four-form potential is
\begin{align}
    \C4 &= \frac{r^4}{L^4} \r^2 \sin\q \, d t \wedge d \r \wedge d \q \wedge d \vf
    \nonumber \\ &\phantom{=}
    - \frac{1}{8} L^4 \le[12 \y^1 - 8 \sin(2\y^1) + \sin(4 \y^1)\ri] \sin^3 \y^2 \sin^2 \y^3 \sin \y^4 \, d\y^2 \wedge d \y^3 \wedge d\y^4.
\end{align}
The curvature radius \(L\) is related to the string coupling \(g_s\) and Regge slope \(\a'\) by \(L^4 = 4 \pi \a'^2 g_s N\). As mentioned in sec.~\ref{sec:intro}, type IIB supergravity on this background is holographically dual to \(\mathcal{N}=4\) SYM with gauge group \(SU(N)\) and 't Hooft coupling \(\l = 4 \pi g_s N\), in the limits \(N \gg \l \gg 1\).

We consider a probe D3-brane embedded into the \(AdS_5 \times S^5\) background. The bosonic part of the D3-brane action is
\begin{equation}
\label{eq:d3_action}
	S_\mathrm{D3} = - T_\mathrm{D3} \int_{\Xi} d^4 \xi \sqrt{-\det\le(g_{ab}+F_{ab}\ri)} + T_\mathrm{D3} \int_{\Xi} P[\C4],
\end{equation}
where the D3-brane tension is \(T_\mathrm{D3} = \le(8 \pi^3 \a'^2 g_s \ri)^{-1}\), \(\xi\) denotes the coordinates on the D3-brane worldvolume \(\Xi\), \(g\) is the pullback of the metric onto \(\Xi\), \(P[\C4]\) is the pullback of \(\C4\) onto \(\Xi\), and \(F =d A\) is the field strength of the \(U(1)\) gauge field $A$ living on the brane. In a D-brane action, $F$ conventionally appears with a coefficient \(2\pi\a'\). We have eliminated this coefficient by re-scaling \(A\) such that our $F$ is dimensionless.

We find it convenient to take the coordinates on the probe D3-brane to be \(\xi = (t,r,\q,\vf)\). For the worldvolume scalar fields we make the ansatz that \(\r\) depends only on \(r\), \(\r=\r(r)\), while the $S^5$ angles are constants, \(\y^I = \y^I_0\). We also assume that the only non-zero components of the gauge field are \(F_{tr}(r)\) and \(F_{\q\vf}(r,\q) \equiv f_{\q\vf}(r)\sin\q\). Evaluated on this ansatz, and after integration over \(\q\) and \(\vf\), the action in eq.~\eqref{eq:d3_action} becomes
\begin{equation}
\label{eq:action_background_ansatz}
    S_\mathrm{D3} = \frac{4\pi T_\mathrm{D3}}{L^4} \int_{\Xi} dt \, dr \le[
        r^4 \r^2 \r'
        - \sqrt{
            \le(L^4 f_{\q\f}^2 + r^4 \r^4 \ri)
            \le(L^4 - L^4 F_{tr}^2 + r^4 \r'^2 \ri)
        }
    \ri].
\end{equation}

The equations of motion that follow from the action in eq.~\eqref{eq:action_background_ansatz} have well-known solutions dual to dyonic BPS solitons on the Coulomb branch~\cite{Schwarz:2014rxa,Schwarz:2014zsa},
\begin{equation}
\label{eq:monopole_solution}
    \r(r) = \frac{L^2 \k}{L v - r},
    \quad
    F_{tr} = \cos \chi,
    \quad
    f_{\q\vf} = L^2 \k \sin\chi,
\end{equation}
with integration constants \(\k\), \(v\), and \(\chi\). We take \(\k > 0\), and hence \(v > 0\) to ensure that \(\r(r)>0\).\footnote{The opposite case of \(\kappa < 0\) leads to a D3-brane with a spike that reaches the boundary of \(AdS_5\). Properties of this solution, including QNMs, were studied in refs.~\cite{Kumar:2016jxy,Kumar:2017vjv,Evans:2019pcs}} If we invert \(\r(r)\), we obtain \(r(\r)\) written in eq.~\eqref{eq:r_rho} with \(\Delta = L v\). The constant $v$ is related to the Coulomb branch adjoint VEV $\varphi$ in ${\cal N}=4$ SYM as $\varphi=\sqrt{N} v/2\pi L$.

For the solution in eq.~\eqref{eq:monopole_solution}, the induced metric on the probe D3-brane is
\begin{equation}
    d s_\Xi^2 = g_{ab} d \xi^a d \xi^b = \frac{r^2}{L^2} \le[
        - d t^2 + \le(\frac{L^4}{r^4} + \frac{L^4\k^2}{(Lv-r)^4} \ri) d r^2
        +\frac{L^4 \k^2}{(Lv -r)^2} \le(  d \q^2 + \sin^2 \q \, d \vf^2 \ri)
    \ri].
\end{equation}
The isometries of this metric, consisting of translations in \(t\) and rotations in \((\q,\vf)\), form a subgroup of the bosonic symmetries preserved by the soliton. Near the Poincar\'e horizon, \(r=0\), the induced metric takes the form of a warped product of \(AdS_2\) with \(S^2\),
\begin{equation}
\label{eq:nearhorizon}
    d s_\Xi^2 \stackrel{r \to 0}{\approx}  - \frac{r^2}{L^2} d t^2 + \frac{L^2}{r^2} d r^2 + \frac{\k^2}{v^2} r^2  \le(  d \q^2 + \sin^2 \q \, d \vf^2 \ri).
\end{equation}
This is already somewhat similar, though not identical, to an extremal black hole, whose near-horizon geometry is $AdS_2 \times S^2$. As for extremal black holes, the $AdS_2$ ``throat'' boundary conditions at $r\to0$ will in part determine the QNMs, which here fall into the Poincar\'e horizon rather than into a black hole event horizon.

The non-zero electric and magnetic fields in eq.~\eqref{eq:action_background_ansatz} imply that the D3-brane contains dissolved fundamental string and D1-brane charge, respectively. To find the string charge, we first compute the canonical momentum conjugate to \(A_t\), which is
\begin{equation}
    \frac{\d S}{\d F_{tr}} = \k L^2 T_\mathrm{D3} \cos \chi \sin\q.
\end{equation}
By Gauss' law, the string charge $p$ is given by the integral over an \(S^2\) centred on \(r=0\),
\begin{equation}
\label{eq:F1_charge}
    p = 2 \pi \a' \int d\q \, d \vf \, \frac{\d S}{\d F_{tr}} = 8 \pi^2 \k \a' L^2 T_\mathrm{D3} \cos \chi = \frac{4 \k N}{\sqrt{\l}} \cos \chi,
\end{equation}
where the factor of \(2\pi\a'\) arises due to our unconventional normalisation of the field strength $F$. The dissolved D1-brane charge $q$ is proportional to the integral of the magnetic field over an \(S^2\) enclosing \(r=0\),
\begin{equation}
\label{eq:D1_charge}
    q = \frac{1}{4 \pi^2 \a'} \int d \q \, d \vf \, F_{\q\vf} = \frac{\k \sqrt{\l}}{\pi} \sin \chi.
\end{equation}
Combining eqs.~\eqref{eq:F1_charge} and~\eqref{eq:D1_charge}, we find that \(\k\) is given in terms of the charges \(p\) and \(q\) by
\begin{equation}
\label{eq:kappa_charges}
    \k = \sqrt{ \frac{\l p^2}{16 N^2} + \frac{\pi^2 q^2}{\l}}.
\end{equation}

The mass of the soliton is, with $S_{\mathrm{D3}} = \int dt dr\mathcal{L}_\mathrm{D3}$~\cite{Schwarz:2014rxa} ,
\begin{equation}
    M = \int d r \, d \q \, d \vf \,  \le[
        F_{tr} \frac{\d S_\mathrm{D3}}{\d F_{tr}} - \mathcal{L}_\mathrm{D3}
    \ri]
    = 4 \pi L^3 \k v T_\mathrm{D3} = \frac{v}{L} \sqrt{ \frac{\l}{4 \pi^2} p^2 + \frac{4 N^2}{\l} q^2 },
    \label{eq:monopole_mass}
\end{equation}
where in the final equality we used eq.~\eqref{eq:kappa_charges} to rewrite \(\kappa\) in terms of \(p\) and \(q\). Since \(M\) is proportional to the charge \(\k\), which is in turn proportional to the radius \(\r_0\) of the soliton, we have that \(M \propto \r_0\), as mentioned in sec.~\ref{sec:intro}. We also note that, recalling the tension of a \((p,q)\)-string \(T_{(p,q)} = (2\pi \a')^{-1} \sqrt{p^2 + q^2/g_s^2}\)~\cite{Becker:2007zj} and using the identification \(\Delta = Lv\) made beneath eq.~\eqref{eq:monopole_solution}, we find \(M = T_{(p,q)}\Delta \), precisely as expected for a \((p,q)\)-string stretched between branes separated by a distance \(\Delta\). In terms of the field theory VEV and gauge couplings, the soliton mass is 
\begin{equation}
M=\varphi \sqrt{(g_{\rm YM} p)^2 + \left(\frac{4\pi q}{g_{\rm YM}}\right)^2}\,,
\end{equation}  
as expected from $SL(2,{\bf Z})$ duality.

\section{Fluctuation Equations}
\label{sec:fluctuation_equations}

\subsection{Action and Equations of Motion}

In this section we consider linearised fluctuations about the D3-brane solutions described in the previous section. For simplicity, we restrict to the purely electrically charged case,\footnote{Worldvolume $SL(2,\mathbb{R})$ transformations can convert the electrically charged solution into magnetically- or dyonically-charged solutions. The linearised fluctuations of those solutions are isomorphic to ours. As a result, those cases will also support QNMs, with spectra isomorphic to ours.} \(\chi=0\), and consider only bosonic fluctuations. We parameterise the fluctuations of the \(AdS_5\) scalar field \(\rho\) and the gauge field \(A\) as
\begin{equation}
\label{eq:fluctuation_definition}
    \r(r) \to \r(r) + \frac{L v}{\k (L v -r)^2} \f(t,r,\q,\vf),
    \qquad
    A_c(t) \to A_c(t) + a_c(t, r,\q,\vf),
\end{equation}
where the normalisation of the fluctuation \(\f\) was chosen to simplify later equations. For the scalar fields in the \(S^5\) directions, it is convenient to decompose the fluctuations in terms of vielbeins \(e^I_A\) on the unit round \(S^5\),
\begin{equation} \label{eq:fluctuation_definition_S5}
    \y^I_0 \to \y^I_0 + e^I_A \, Z^A(t,r,\q,\vf).
\end{equation}
Following ref.~\cite{Faraggi:2011bb}, we will make use of the \(SO(3)\) symmetry corresponding to rotations in the \((\q,\vf)\) directions to decompose the fluctuations into spherical harmonics,
\begin{align}
    \f(t,r,\q,\vf) &= \sum_{l=0}^\infty \sum_{m=-l}^{l} \f_{lm}(t,r) Y^{lm}(\q,\vf),
    \nonumber \\
    a_\a(t,r,\q,\vf) &= \sum_{l=0}^\infty \sum_{m=-l}^{l} a^{lm}_\a(t,r)  Y^{lm}(\q,\vf),
    \nonumber \\
    a_i(t,r,\q,\vf) &= \sum_{l=1}^\infty \sum_{m=-l}^{l} \le[\b_{lm}(t,r) Y_i^{lm}(\q,\vf) + \frac{L v}{L v -r} b_{lm}(t,r) \hat Y_i^{lm}(\q,\vf) \ri],
    \label{eq:spherical_harmonic_expansion}
    \\
    Z^A(t,r,\q,\vf) &= \sum_{l=0}^\infty \sum_{m=-l}^l Z^A_{lm}(t,r) Y^{lm}(\q,\vf),
    \nonumber
\end{align}
where \(\a\in\{t,r\}\) and \(i\in\{\q,\vf\}\), the vector spherical harmonics are \(Y_i^{lm}(\q,\vf) = \p_i Y^{lm}(\q,\vf) \) and \(\hat{Y}_i^{lm}(\q,\vf) = \frac{1}{\sqrt{l(l+1)}} \e_i{}^j \p_j Y^{lm} \), and the normalisation  of \(b_{lm}(t,r)\) has been chosen for later convenience. Note that the \(a_i\) fluctuations begin at \(l=1\) rather than $l=0$. The coefficients \(\b_{lm}\) may be eliminated by a gauge transformation of the form~\cite{Faraggi:2011bb}
\begin{equation}
    A_c(t,r,\q,\vf) \to A_c(t,r,\q,\vf) - \p_c  \sum_{l=1}^\infty \sum_{m=-l}^{l} \b_{lm}(t,r) Y^{lm}(\q,\vf).
\end{equation} 
We will work in the gauge \(\b_{lm}=0\) throughout.

We obtain an action for the fluctuations by expanding the action eq.~\eqref{eq:d3_action} to quadratic order in \(\f\), \(a_a\), and \(Z^A\), plugging in the spherical harmonic decompositions of eq.~\eqref{eq:spherical_harmonic_expansion}, and integrating over \(\q\) and \(\vf\). We find that many fluctuations decouple from one another, such that the part of the action quadratic in fluctuations splits into three terms,
\begin{subequations}
\label{eq:fluctuation_action}
\begin{equation}
    S = S_1 + S_2 + S_3.
    \label{eq:fluctuation_action_total}
\end{equation}
To express $S_{1}$, $S_2$, and $S_{3}$ compactly, we introduce a dimensionless time coordinate, \(\t \equiv v t/L\), and a dimensionless radial coordinate \(x \equiv r/Lv\), valued in the range \(x \in [0,1]\) on the D3-brane worldvolume. The first term in eq.~\eqref{eq:fluctuation_action}, \(S_1\), is an action for the \(AdS_5\) scalar fluctuation \(\f\) and the gauge field fluctuations \(a_{\t}\) and \(a_{x}\),
\begin{align}
    S_1 &= \frac{T_\mathrm{D3}}{2 \k } \sum_{l=0}^\infty \sum_{m=-l}^l \int d \t \, d x \biggl[
        \frac{(\dot{\f}_{lm})^2}{(1-x)^4} - 2 \f_{lm}' \cF_{\t x}^{l,-m}
        + \frac{2l(l+1)}{ (1-x)^2} \f_{lm} a_\t^{l,-m}
        + \frac{f(x)}{x^4} (\cF_{\t x}^{lm})^2
        \nonumber \\ &\phantom{= \frac{T_\mathrm{D3}}{2 \k } \sum_{l=0}^\infty \sum_{m=-l}^l \int d \t d x \biggl[}
        - l(l+1) (1-x)^2  (a_x^{lm})^2
        + l(l+1) \frac{f(x)}{x^4(1-x)^2} (a_\t^{lm})^2
    \biggr],
    \label{eq:fluctuation_action_electric}
\end{align}
\begin{equation}
f(x) \equiv \k^2 x^4 + (1-x)^4,
\end{equation}
where dots denote derivatives with respect to \(\t\), primes denote derivatives with respect to \(x\), and \(\cF_{\t x} \equiv \dot{a}_x - a_\t'\). In this and all subsequent expressions, any term of the form \((X_{lm})^2\) means \(X_{lm} X_{l,-m}\), where \(X\) is any fluctuation or its derivatives. The second term in eq.~\eqref{eq:fluctuation_action}, \(S_2\), depends only on the magnetic field fluctuations, \(b_{lm}\),
\begin{equation}
    S_2 = \frac{T_\mathrm{D3}}{2 \k}  \sum_{l=1}^\infty \sum_{m=-l}^l (-1)^m \int d \t \, d x\, \biggl[ \frac{f(x)}{x^4(1-x)^4} (\dot{b}_{lm})^2
    -\le(b'_{lm} - \frac{b_{lm}}{1-x}\ri)^2 - \frac{l(l+1)}{(1-x)^2} (b_{lm})^2 \biggr].
    \label{eq:flucutation_action_magnetic}
\end{equation}
The third term in eq.~\eqref{eq:fluctuation_action}, $S_3$, depends only on the \(S^5\) fluctuations \(Z^A\),
\begin{equation}
    S_3 = \frac{L^2 \k T_\mathrm{D3}}{2 v^2} \sum_{A=1}^5 \sum_{l=0}^\infty \sum_{m=-l}^l (-1)^m \int d \t \, d x \biggl[
        \frac{f(x)}{x^4(1-x)^4} (\dot{Z}_{lm}^A)^2
        - ({Z_{lm}^{A}}')^2  - \frac{l(l+1)}{(1-x)^2} (Z_{lm}^A)^2
    \biggr].
    \label{eq:flucutation_action_S5}
\end{equation}
\end{subequations}

The equations of motion may be obtained straightforwardly by varying the action eq.~\eqref{eq:fluctuation_action} with respect to the fluctuations. Varying \(S_1\) in eq.~\eqref{eq:fluctuation_action_electric}, we obtain three coupled equations of motion for \(\f\), \(a_\t\), and \(a_x\),
\begin{subequations} \label{eq:electric_sector_eoms}
\begin{align}
    \ddot{\f}_{lm} - (1-x)^4 {\cF_{\t x}^{lm}}' - (1-x)^2 l(l+1) a_\t^{lm} &= 0,
    \label{eq:general_l_phi_eom}
    \\
    \p_x\le(
        \f_{lm}' -  \frac{f(x)}{x^4} \cF_{\t x}^{lm}
    \ri) - \frac{l(l+1)}{(1-x)^2} \le(\f + \frac{f(x)}{x^4} a_\t^{lm} \ri) &=0,
    \label{eq:general_l_at_eom}
    \\
    \p_\t \le(
        \f_{lm}' - \frac{f(x)}{x^4} \cF_{\t x}^{lm} 
    \ri) - l(l+1) (1-x)^2 a_x^{lm} &=0.
    \label{eq:general_l_ax_eom}
\end{align}
\end{subequations}
We can reduce eq.~\eqref{eq:electric_sector_eoms} to two decoupled equations by defining the new dependent variables,
\begin{align} \label{eq:decoupled_fluctuations}
    \F^{lm}_{1} &\equiv (l+1)\f_{lm} + (1-x) \f_{lm}' - \frac{f(x)}{x^4}(1-x) \cF_{\t x}^{lm},
    \nonumber \\ 
    \F^{lm}_{2} &\equiv l \f_{lm} - (1-x) \f_{lm}' + \frac{f(x)}{x^4}(1-x) \cF_{\t x}^{lm}.
\end{align}
It will be convenient to Fourier transform with respect to \(\t\). Since the action is invariant under constant shifts in \(\t\), different Fourier modes decouple. We may therefore consider only a single mode of each fluctuation, with dimensionless frequency, \(\wb\), such as \(\F_1^{lm}(\t,x) = e^{-i \wb \t} \, \F_1^{lm}(x)\). From eqs.~\eqref{eq:electric_sector_eoms} and~\eqref{eq:decoupled_fluctuations} we thus find
\begin{align} \label{eq:higher_l_Phi_eom}
    {\F_1^{lm}}'' + \frac{4(1-x)^3}{x f(x)} {\F_1^{lm}}' + \le[ \frac{\wb^2 f(x)}{x^4 (1-x)^4} - \frac{l(l+1)}{(1-x)^2} - \frac{4 l (1-x)^2}{x f(x)} \ri] \F_{1}^{lm}
    &= 0,
    \nonumber \\
    {\F_{2}^{lm}}'' + \frac{4(1-x)^3}{x f(x)} {\F_{2}^{lm}}' + \le[ \frac{\wb^2 f(x)}{x^4 (1-x)^4} - \frac{l(l+1)}{(1-x)^2} + \frac{4(l+1)(1-x)^2}{x f(x)}  \ri] \F_{2}^{lm}
    &= 0.
\end{align}

Care must be taken when \(l=0\) (and therefore \(m=0\)). When $l=0$, eq.~\eqref{eq:general_l_ax_eom} implies \(\f'_{00} - \frac{f(x)}{x^4} \cF_{\t x}^{00} = 0\) for any non-zero \(\wb\). Substituting this into eq.~\eqref{eq:decoupled_fluctuations} with \(l=0\), we find that this implies \(\F_2^{00} = 0\). As a result, at \(l=0\) the only non-trivial fluctuation in the \((\f,a_{\t},a_x)\) channel is \(\F_1^{00}\). For any \(l \neq 0\) both \(\F_1^{lm}\) and \(\F_2^{lm}\) are non-trivial.

Varying \(S_2\) in eq.~\eqref{eq:flucutation_action_magnetic} with respect to \(b_{lm}\) and performing a Fourier transform with respect to \(\t\), we obtain the equation of motion for the magnetic field fluctuations,
\begin{equation}
\label{eq:b_EOM}
    b_{lm}'' + \le[ \frac{\wb^2 f(x)}{x^4(1-x)^4} - \frac{l(l+1)}{(1-x)^2}\ri] b_{lm} = 0.
\end{equation}

Finally, varying \(S_3\) in eq.~\eqref{eq:flucutation_action_S5} with respect to \(Z^A_{lm}\), we obtain the equation of motion for the scalar field fluctuations on the \(S^5\). From now on we will drop the superscript on \(Z^A_{lm}\), since the equation of motion does not depend on the value of the index \(A\). The resulting equation of motion is the same as eq.~\eqref{eq:b_EOM}, since the \(l>1\) terms in \(S_2\) and \(S_3\) are equivalent up to a boundary term under interchange of \(b_{lm} \leftrightarrow Z_{lm}\),
\begin{equation}
\label{eq:S5_EOM}
    Z_{lm}'' + \le[ \frac{\wb^2 f(x)}{x^4(1-x)^4} - \frac{l(l+1)}{(1-x)^2}\ri] Z_{lm} = 0.
\end{equation}
However, note that \(Z_{lm}\) has \(l\geq0\), whereas \(b_{lm}\) has \(l \geq 1\).

We thus have four second-order equations of motion for the fluctuations \(\F^{lm}_1\), \(\F^{lm}_2\), \(b_{lm}\), and \(Z_{lm}\). At both the Poincar\'e horizon (\(x=0\)) and spatial infinity (\(x=1\)), we find that each equation of motion admits independent ingoing and outgoing solutions. For example, near \(x=0\) we find that solutions to the equation of motion eq.~\eqref{eq:S5_EOM} take the form
\begin{equation}
\label{eq:ZPoincaresol}
    Z_{lm} = x \, c_+ \, g_+(x)  e^{i \wb/x} + x \, c_- \, g_-(x) e^{-i \wb/x},
\end{equation}
where \(c_\pm\) are constants, and the functions \(g_\pm(x)\) are regular at \(x=0\) and normalised such that \(g_\pm(0) = 1\). The form in eq.~\eqref{eq:ZPoincaresol} is that expected for the warped product of $AdS_2$ and $S^2$ in eq.~\eqref{eq:nearhorizon}. Ingoing boundary conditions correspond to the choice \(c_+ = 0\), while outgoing boundary conditions correspond to \(c_- = 0\). Similarly, near \(x=1\) we find
\begin{equation}
\label{eq:Zasymptotic}
    Z_{lm} = (1-x) \, d_+ \, h_+(x) e^{i \k \wb/(1-x)} + (1-x)  \, d_- \, h_-(x)  e^{-i \k \wb/(1-x)},
\end{equation}
where \(d_\pm\) are constants, and the functions \(h_\pm(x)\) are regular at \(x=1\), with \(h_\pm(1)=1\). The form in eq.~\eqref{eq:Zasymptotic} is that expected for $(3+1)$-dimensional flat space. Ingoing boundary conditions correspond to \(d_+ = 0\), and outgoing to \(d_- =0 \). We define QNMs as solutions of the fluctuation equations satisfying outgoing boundary conditions at both \(x=0\) and \(x=1\), i.e. \(c_- = 0\) and \(d_- = 0\). Such solutions exist only for certain frequencies \(\wb\). 

Physically, the QNM frequencies are poles of the Green's functions for the fluctuation equations that satisfy outgoing boundary conditions at \(x=0\) and \(x=1\). These Green's functions may also have branch point singularities. Indeed, as described below, we find evidence for a branch point at \(\wb = 0\) in all fluctuations channels and all values of \(l\).

\subsection{Isospectrality}
\label{sec:isospectrality}

The QNM spectra of \(b_{lm}\) and \(Z_{lm}\) are the same, since their equations of motion, eqs.~\eqref{eq:b_EOM} and~\eqref{eq:S5_EOM}, are identical. Less trivially, we find that they also have the same spectrum as \(\F_1^{l-1,m}\) and \(\F_2^{l+1,m}\). In other words, all four fluctuation channels are isospectral. To prove this, we will compute the Schr\"odinger potentials for the fluctuations, as follows.

Consider a second order linear differential equation of the form
\begin{equation}
\label{eq:2nd_order_ode}
    y''(x) + P(x) y'(x) + \le[\wb^2 Q(x) - R(x) \ri] y(x) = 0,
\end{equation}
which we wish to put into Schr\"odinger form
\begin{equation} \label{eq:general_schrodinger}
    \frac{d^2 \y(x_*)}{d x_*^2} + \le[\wb^2 - V(x_*) \ri] \y(x_*) = 0,
\end{equation}
with some Schr\"odinger potential, \(V(x_*)\). To do so, we define the tortoise coordinate \(x_*\) via
\begin{equation} \label{eq:tortoise_coordinate}
    \frac{d x_*}{d x} = \sqrt{Q(x)}.
\end{equation}
This eliminates the coefficient of \(\wb^2\) inside the square brackets in eq.~\eqref{eq:2nd_order_ode}, so that the differential equation eq.~\eqref{eq:2nd_order_ode} becomes
\begin{equation} \label{eq:2nd_order_ode_tortoise}
    \frac{d^2 y(x_*)}{d x_*^2} + \frac{Q'(x) + 2 P(x)Q(x)}{2 Q(x)^{3/2}} \frac{d y(x_*)}{d x_*} + \le[ \wb^2 - \frac{R(x)}{Q(x)} \ri] y(x_*)= 0.
\end{equation}
In this expression, \(x\) should be interpreted as an implicit function of \(x_*\), determined from eq.~\eqref{eq:tortoise_coordinate}. We may then eliminate the first derivative term by defining a new dependent variable \(\y(x_*) = e^{h(x_*)} y(x_*)\), where \(h\) satisfies
\begin{equation}
    \frac{d h(x_*)}{d x_*} = \frac{Q'(x) + 2 P(x) Q(x)}{4 Q(x)^{3/2}}.
\end{equation}
This puts eq.~\eqref{eq:2nd_order_ode_tortoise} into the Schr\"odinger form of eq.~\eqref{eq:general_schrodinger}, with Schr\"odinger potential
\begin{equation}
\label{eq:schrodinger_potential_general}
    V(x_*) = \frac{R(x)}{Q(x)} + \frac{Q''(x)}{4 Q(x)^2} - \frac{5 Q'(x)^2}{16 Q(x)^3} + \frac{2 P'(x) + P(x)^2}{4 Q(x)},
\end{equation}
where, again, \(x\) should be interpreted as a function of \(x_*\) obtained from eq.~\eqref{eq:tortoise_coordinate}.

Each of our four fluctuations satisfies an equation of motion of the form in eq.~\eqref{eq:2nd_order_ode}. The function \(Q(x)\) is the same for all four cases, \(Q(x) = f(x)/x^4(1-x)^4\). Solving eq.~\eqref{eq:tortoise_coordinate}, we obtain the tortoise coordinate
\begin{equation}
\label{eq:tortoise_solution}
    x_* = c + \frac{\k x}{1-x} \, {}_2 F_1 \le( - \frac{1}{2}, - \frac{1}{4}; \frac{3}{4}; - \frac{(1-x)^4}{\k^2 x^4} \ri),
\end{equation}
where \(c\) is an integration constant. Under this coordinate transformation, the Poincar\'e horizon at \(x=0\) is mapped to \(x_* = - \infty\), while spatial infinity (\(\r \to \infty\)) at \(x=1\) is mapped to \(x_* = \infty\). In general we cannot solve eq.~\eqref{eq:tortoise_solution} for \(x\) as a function of \(x_*\). However, we can determine the asymptotic behaviour of \(x\) near \(x_* = \pm \infty\) by expanding the hypergeometric function in eq.~\eqref{eq:tortoise_solution} near \(x=0\) and \(x=1\), obtaining
\begin{equation} 
\label{eq:tortoise_asymptotics}
    x \sim \begin{cases}
        - \dfrac{1}{x_*}, & x_* \to - \infty,
        \\[0.5em]   
        1 - \dfrac{\k}{x_*}, & x_* \to  \infty.
    \end{cases}
\end{equation}

The various fluctuation channels have different functions \(P(x)\) and \(R(x)\), and therefore different Schr\"odinger potentials from eq.~\eqref{eq:schrodinger_potential_general}. The potentials for \(\F_1^{lm}\), \(\F_2^{lm}\), and \(Z_{lm}\) are
\begin{align}
    V_{\F_1,l}(x_*) &= \frac{l (l+1) x^4 (1-x)^2}{f(x)} - \frac{\le[5 - 4 (l+1) x\ri] x^2 (1-x)^6}{f^2(x)} + \frac{7 x^2 (1-x)^{10}}{f^3(x)},
    \nonumber
    \\
    V_{\F_2,l}(x_*) &= \frac{l (l+1) x^4 (1-x)^2}{f(x)} - \frac{\le(5 +  4 l x \ri) x^2 (1-x)^6}{f^2(x)} + \frac{7 x^2 (1-x)^{10}}{f^3(x)},
    \label{eq:schrodinger_potentials}
    \\
    V_{Z,l}(x_*) &= \frac{l(l+1) x^4 (1-x)^2}{f(x)} + \frac{5 x^2 (1-x)^6}{f^2(x)} - \frac{5 x^2 (1-x)^{10}}{f^3(x)},
    \nonumber
\end{align}
respectively. Since the equation of motion for \(b_{lm}\) is the same as that for \(Z_{lm}\), its Schr\"odinger potential is \(V_{b,l}(x_*) = V_{Z,l}(x_*)\).

Using eq.~\eqref{eq:tortoise_asymptotics}, we can determine the leading order asymptotics of the potentials as \(x_* \to \pm \infty\). We find that \(V_{\F_1,l}\) and \(V_{\F_2,l}\) exhibit the same behaviour in both limits,
\begin{equation}
    V_{\F_{1,2},l}(x_*) \sim \begin{cases}
        \dfrac{2}{x_*}, & x_* \to - \infty,
        \\[0.5em]
        \dfrac{l(l+1)}{x_*^2}, & x_* \to \infty,
    \end{cases}
\end{equation}
while the potential for the \(S^5\) fluctuations has the leading order asymptotics
\begin{equation}
\label{eq:schrodinger_Z_asymptotics}
    V_{Z,l}(x_*) \sim \begin{cases}
            \dfrac{l(l+1)}{x_*^4}, & x_* \to - \infty,
        \\[0.5em]
        \dfrac{l(l+1)}{x_*^2}, & x_* \to \infty.
    \end{cases}
\end{equation}

We can now prove the isospectrality mentioned at the start of this section. The method we use is standard, see for example the review ref.~\cite{Berti:2009kk} and references therein, and works as follows. Consider a pair of fields \(\y_\pm\) satisfying the Schr\"odinger equations
\begin{equation}
\label{eq:psi_plus_minus_schrodinger}
    \frac{d^2 \y_\pm(x_*)}{d x_*^2} + \le[ \wb^2 - V_\pm(x_*)\ri] \y_\pm(x_*) = 0,
\end{equation}
where the potentials \(V_\pm(x_*)\) may be written in terms of a ``superpotential'' \(W(x_*)\) as,
\begin{equation}
\label{eq:superpotential}
    V_\pm(x_*) = W^2(x_*) \mp \frac{d W(x_*)}{d x_*} + \Omega^2,
\end{equation}
where \(\Omega\) is a constant. It is then straightforward to show that, given \(\y_{\pm}(x_*)\) that satisfies eq.~\eqref{eq:psi_plus_minus_schrodinger}, the combination $\pm W(x_*) \y_{\pm}(x_*) + \frac{d \y_{\pm}(x_*)}{d x_*}$ solves the equation for $\psi_{\mp}(x_*)$. In other words, given one solution $\psi_{\pm}(x_*)$, if we can find $W(x_*)$ obeying eq.~\eqref{eq:superpotential} then we can construct the other solution as 
\begin{equation}
\label{eq:isospectrality_transformation}
    \y_\mp(x_*)= \pm W(x_*) \y_\pm(x_*) + \frac{d \y_\pm(x_*)}{d x_*}.
\end{equation}
It is also straightforward to show that if \(\y_{\pm}(x_*)\) satisfies outgoing boundary conditions at \(x_* \to \pm \infty\), then so does \(\y_\mp(x_*)\) in eq.~\eqref{eq:isospectrality_transformation}. Hence, \(\y_\pm(x_*)\) are isospectral --- they share the same spectrum of QNMs.

Our task is thus to show that \(V_{\F_1,l-1}\) and \(V_{\F_2,l+1}\) may both be paired with \(V_{Z,l}\) through relations of the form in eq.~\eqref{eq:superpotential}. Using the potentials in eq.~\eqref{eq:schrodinger_potentials}, and the definition of the tortoise coordinate in eq.~\eqref{eq:tortoise_coordinate} with \(Q(x) = f(x)/x^4(1-x)^4\), we indeed find
\begin{subequations}
\begin{equation}
    V_{Z,l}(x_*) = W_{1,l}^2(x_*) - \frac{d W_{1,l}(x_*)}{d x_*},
    \quad
    V_{\F_1,l-1}(x_*) = W_{1,l}^2(x_*) + \frac{d W_{1,l}(x_*)}{d x_*},
\end{equation}
\begin{equation}
    V_{Z,l}(x_*) = W_{2,l}^2(x_*) - \frac{d W_{2,l}(x_*)}{d x_*},
    \quad
    V_{\F_2,l+1}(x_*) = W_{2,l}^2(x_*) + \frac{d W_{2,l}(x_*)}{d x_*},
\end{equation}
\end{subequations}
with the superpotentials
\begin{equation} \label{eq:superpotential_results}
    W_{1,l}(x_*) = \frac{l x^2 (1-x)}{f(x)^{1/2}} + \frac{x(1-x)^5}{f(x)^{3/2}},
    \quad
    W_{2,l}(x_*) =  - \frac{(l+1) x^2 (1-x)}{f(x)^{1/2}} + \frac{x (1-x)^5}{f(x)^{3/2}}.
\end{equation}
As a result, \(\F_1^{l-1,m}\) and \(\F_2^{l+1,m}\) must share the same QNM spectra as \(Z_{lm}\) and \(b_{lm}\).

The fluctuations can therefore be combined into multiplets \((Z^A_{lm},b_{lm},\F_1^{l-1,m},\F_2^{l+1,m})\), where each component has the same QNM spectrum. Note that since \(b_{lm}\) and \(\F_1^{l-1,m}\) have \(l \geq 1\), the \(l=0\) multiplet is only \((Z^A_{00}, \F_2^{1,m})\).

This isospectrality is presumably due to supersymmetry, i.e. the multiplets we have found are probably the bosonic parts of a multiplet of the supergroup preserved by the D3-brane solution in eq.~\eqref{eq:monopole_solution}. We will leave a detailed analysis of that supergroup, its multiplets, and their relation to our isospectral multiplet to future research.

\section{Quasi-Normal Modes}
\label{sec:results}

In this section we present results for our QNM spectra. Since the fluctuation channels share the same spectrum, as shown in sec.~\ref{sec:isospectrality}, we will consider only the D3-brane fluctuations on the $S^5$,  \(Z_{lm}\), which satisfy the equation of motion eq.~\eqref{eq:S5_EOM}.

At \(l=0\), we can show that all QNM frequencies must depend on \(\k\) as \(\wb \propto 1/\sqrt{\k}\). To do so, we define new variables \(u \equiv \ln \le[\sqrt{\k} \, x/(1-x) \ri]\) and \(\tilde{Z} \equiv \le(e^{u/2} + \sqrt{\k} e^{-u/2}\ri) Z_{00}\), in terms of which eq.~\eqref{eq:S5_EOM} with \(l=0\) becomes the modified Mathieu equation,
\begin{equation}
    \tilde{Z}''(u) - \le[\frac{1}{4} - 2 \k \wb^2 \cosh(2u) \ri] \tilde{Z}(u) = 0.
\end{equation}
Since \(\k\) and \(\wb\) appear only in the combination \(\k \wb^2\), the QNMs must have \(\wb \propto 1/\sqrt{\k}\).

More generally, to determine the QNMs we employ a numerical method, Leaver's matrix method~\cite{Leaver:1990zz,Onozawa:1995vu}. In this method, we write the equation of motion eq.~\eqref{eq:S5_EOM} as an infinite-dimensional matrix equation for a set of coefficients \(\cZ_{lm,I}\),
\begin{equation}
    \sum_{J=0}^\infty C_{IJ}(\wb) \mathcal{Z}_{lm,J} = 0,
\end{equation}
for some matrix \(C_{IJ}(\wb)\). This equation only has non-trivial solutions when \(\det C_{IJ}(\wb) = 0\). The values of \(\wb\) at which this occurs are precisely the QNM frequencies. We can approximately determine these frequencies by truncating the matrix from infinite-dimensional to \(M \times M\) with finite $M$, i.e. restricting \(I,J \leq M-1\), and then solving \(\det C_{IJ}(\wb) = 0\) numerically for \(\wb\). The larger \(M\) is, the better we expect the approximation to be. Further details of this method, including expressions for the components of \(C_{IJ}(\wb)\) and the definition of the coefficients \(\mathcal{Z}_{lm,J}\), are given in appendix~\ref{app:leavers}.

\begin{figure}
    \begin{subfigure}{\textwidth}
    \begin{center}
            \includegraphics{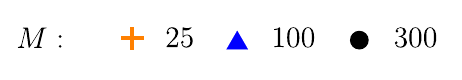}
    \end{center}
    \end{subfigure}
    \begin{subfigure}{0.5\textwidth}
        \includegraphics[width=\textwidth]{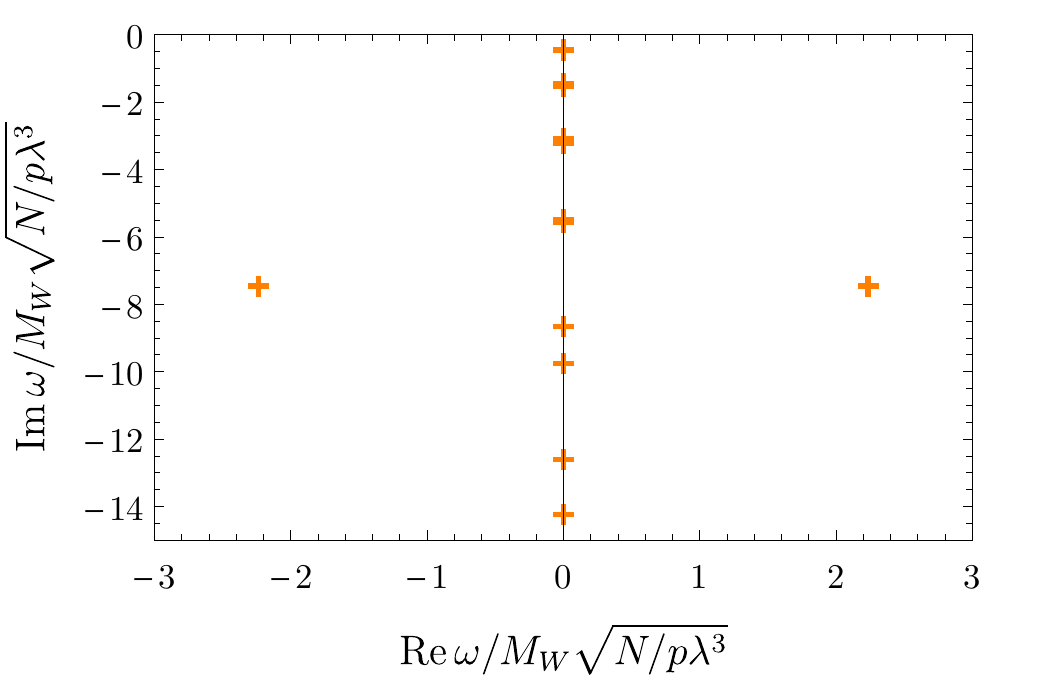}
        \caption{\(M=25\)}
    \end{subfigure}
    \begin{subfigure}{0.5\textwidth}
        \includegraphics[width=\textwidth]{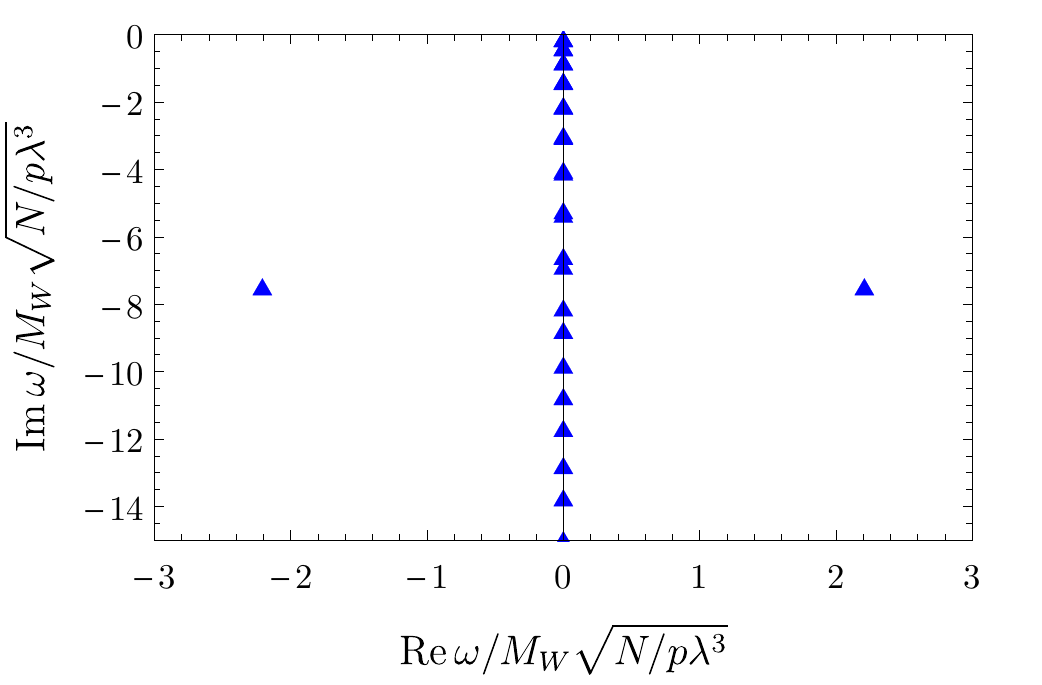}
        \caption{\(M=100\)}
    \end{subfigure}
    \begin{subfigure}{0.5\textwidth}
        \includegraphics[width=\textwidth]{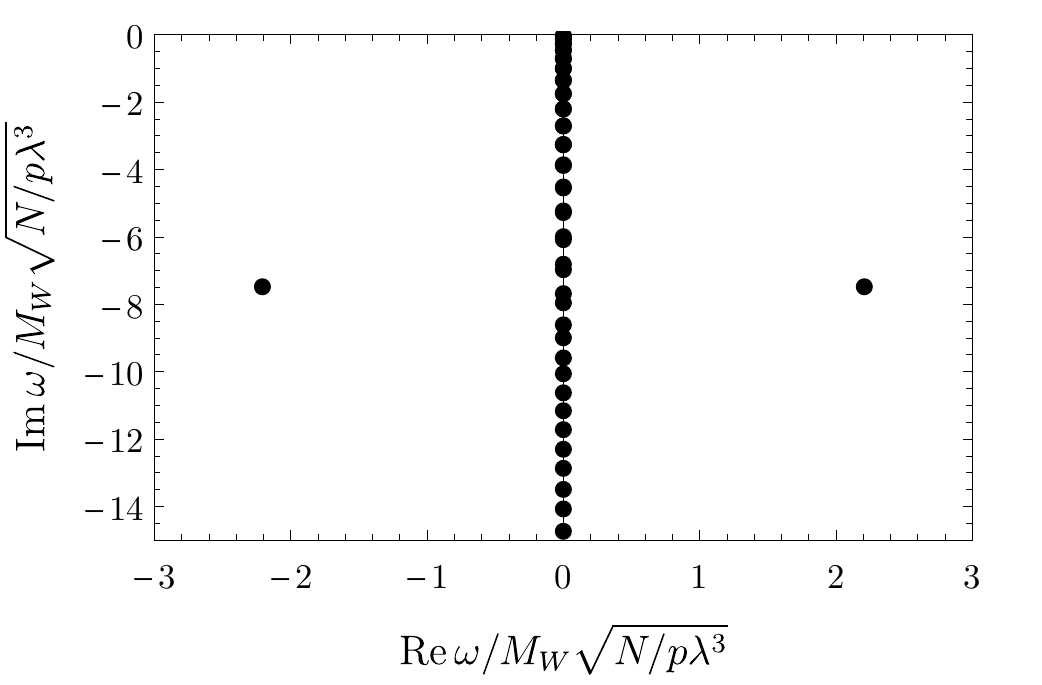}
        \caption{\(M=300\)}
    \end{subfigure}
    \begin{subfigure}{0.5\textwidth}
        \includegraphics[width=\textwidth]{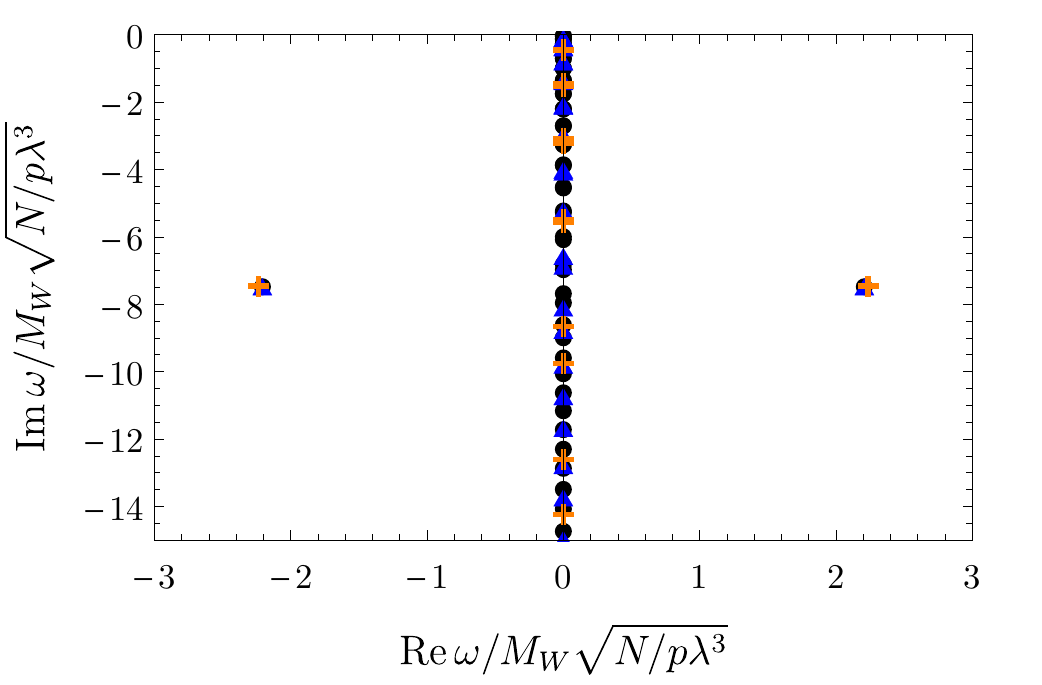}
        \caption{$M=25,100,300$ superimposed}
    \end{subfigure}
    \caption{QNM frequencies at \(l=0\), in units of \(M_W \sqrt{N/p\l^3}\). The frequencies were computed numerically using Leaver's matrix method, detailed in appendix~\ref{app:leavers}. The different symbols and colours correspond to different matrix sizes \(M\): orange crosses to \(M=25\), blue triangles to \(M=100\), and black dots to \(M=300\). \textbf{(a,\,b,\,c):} QNM frequencies computed with these different values of \(M\). For all matrix sizes we observe a pair of modes with equal and opposite \(\Re \w\neq 0\), and \(\Im \w<0\). The frequencies of these QNMs computed with \(M=300\) are given in eq.~\eqref{eq:l0_modes}. We also observe many poles on the negative imaginary axis, which become denser as we increase \(M\), consistent with a branch cut. \textbf{(d):} The results for our three different \(M\) values, superimposed. The isolated modes do not appear to move as we increase \(M\), indicating good convergence of our numerical method.}
    \label{fig:l0_modes}
\end{figure}

Fig.~\ref{fig:l0_modes} shows the QNM frequencies we find at \(l=0\), for matrix sizes \(M = 25\), \(100\), and \(300\), indicated by the orange crosses, blue triangles, and black dots, respectively. For all \(M\), we find a single pair of QNM frequencies with non-zero real parts. Using our highest precision \(M=300\) numerics, we find the dimensionless frequencies of these QNMs to be \( \wb \approx \le(\pm 0.175720 - 0.59531 i\ri)/\sqrt{\k} \). We also define a \textit{dimensionful} frequency \(\w = v \wb/L\), such that \(\wb \t = \w t\). Using eq.~\eqref{eq:F1_charge} with \(\chi=0\) to relate \(\k\) to \(p\), we then find that this isolated pair of modes has dimensionful frequency
\begin{equation}
\label{eq:l0_modes}
    \w \approx \le(\pm 2.20816 - 7.48087 i \ri) M_W \sqrt{\frac{N}{p \l^3}},
\end{equation}
where \(M_W = v \sqrt{\l} / 2 \pi L\) is the W-boson mass.

In addition to these two isolated QNMs, in fig.~\ref{fig:l0_modes} for all \(M\) we find many QNMs on the negative imaginary axis. These modes become denser as we increase \(M\). We expect that these modes arise from a branch cut of the Green's function for eq.~\eqref{eq:S5_EOM} along the negative imaginary axis. In Leaver's method, the branch cut appears as a finite number of isolated poles due to the truncation to finite \(M\)~\cite{Denef:2009yy,Edalati:2010hk,Edalati:2010pn}.

As a check of our numerics, we have confirmed that we obtain all the same QNM frequencies using the spectral method of the \texttt{Mathematica} package \texttt{QNMspectral}~\cite{Jansen:2017oag}.

\begin{figure}
\begin{subfigure}{\textwidth}
    \begin{center}
        \includegraphics{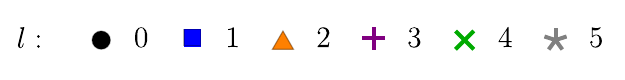}
    \end{center}
\end{subfigure}
    \begin{subfigure}{0.5\textwidth}
        \includegraphics[width=\textwidth]{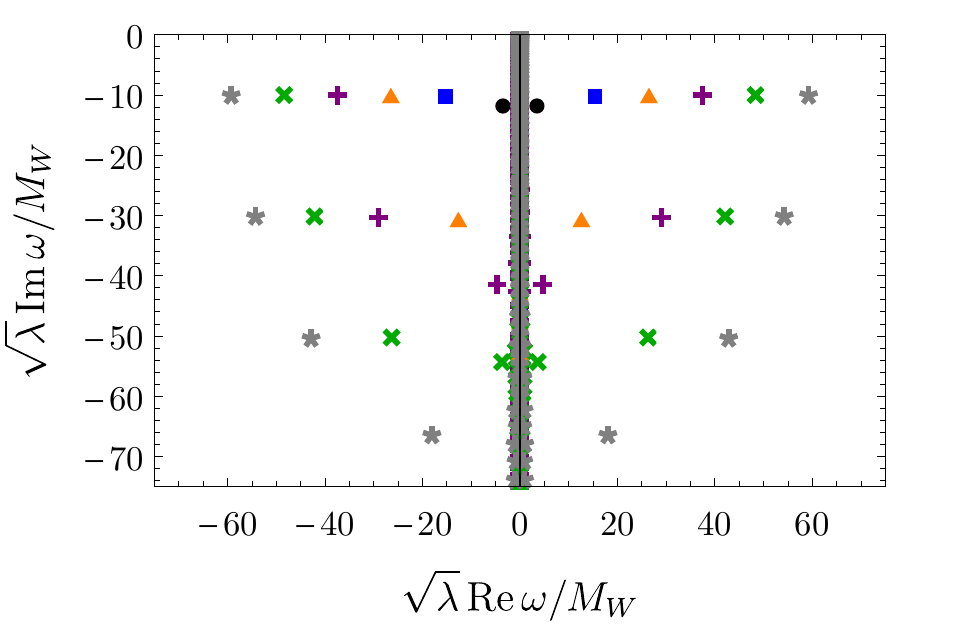}
        \caption{\(\k=1/10\)}
    \end{subfigure}
    \begin{subfigure}{0.5\textwidth}
        \includegraphics[width=\textwidth]{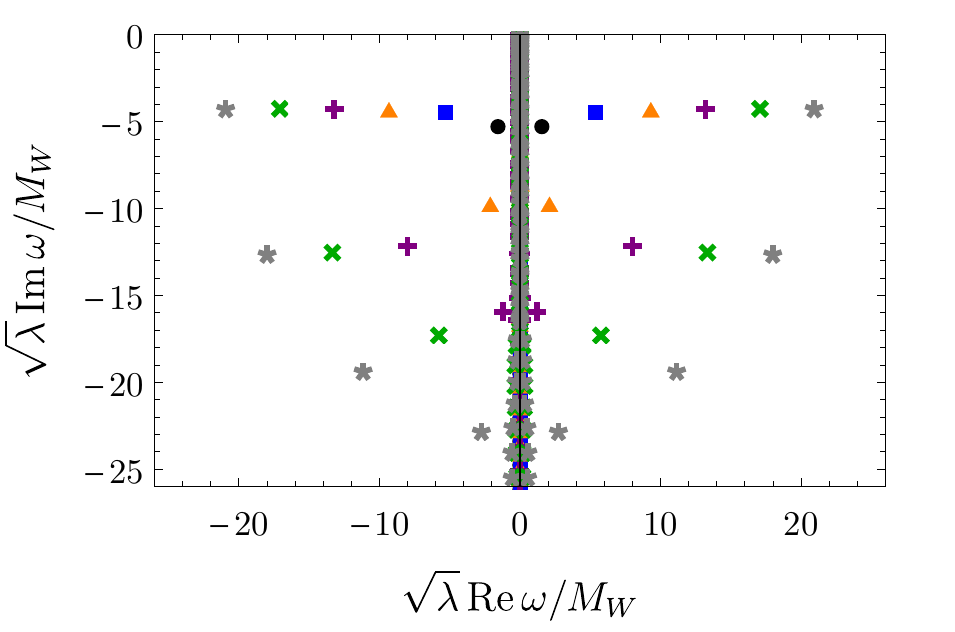}
        \caption{\(\k=1/2\)}
    \end{subfigure}
    \begin{subfigure}{0.5\textwidth}
        \includegraphics[width=\textwidth]{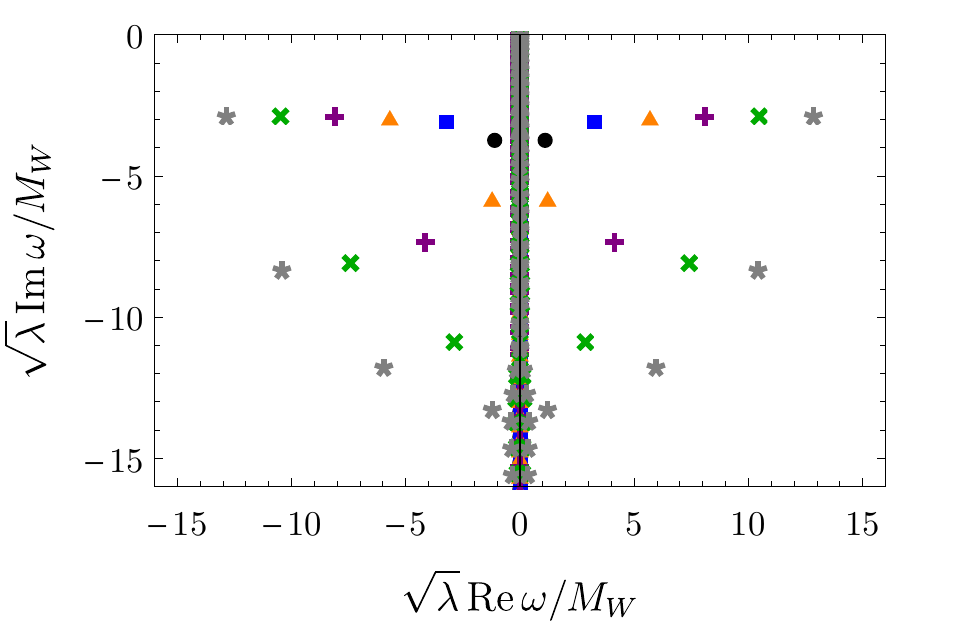}
        \caption{\(\k=1\)}
    \end{subfigure}
    \begin{subfigure}{0.5\textwidth}
        \includegraphics[width=\textwidth]{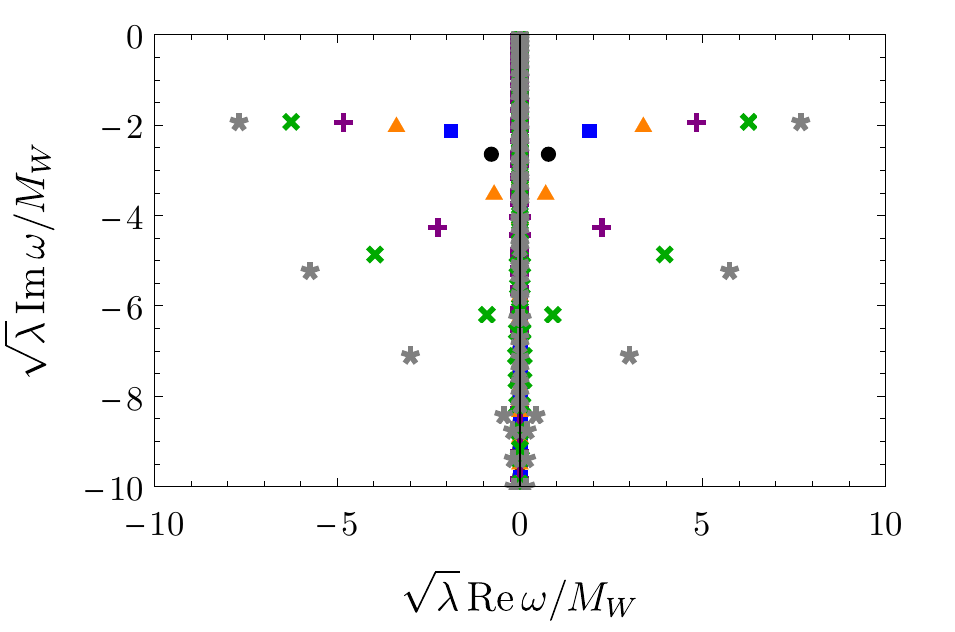}
        \caption{\(\k=2\)}
    \end{subfigure}
    \begin{subfigure}{0.5\textwidth}
        \includegraphics[width=\textwidth]{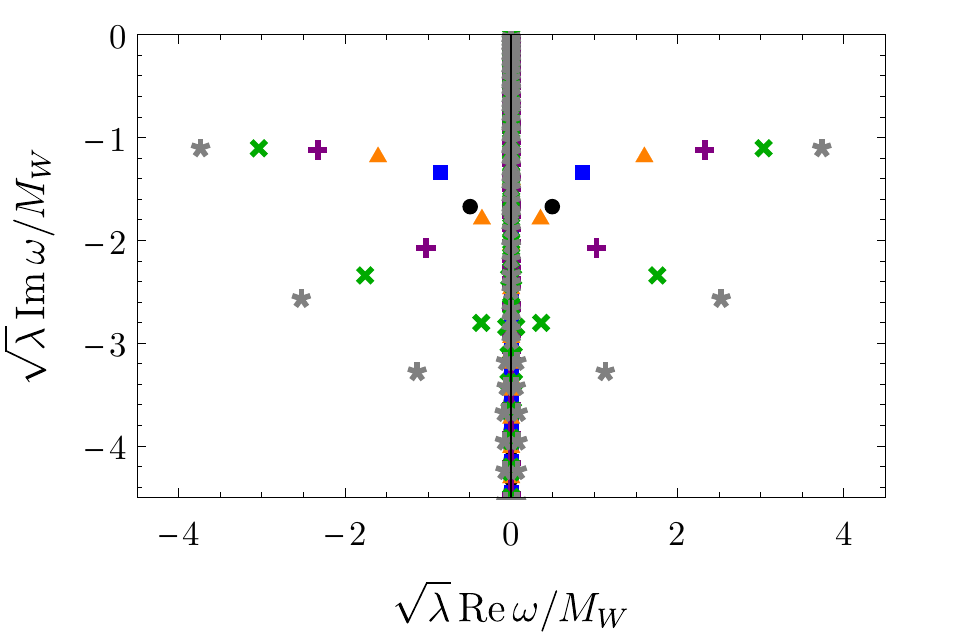}
        \caption{\(\k=5\)}
    \end{subfigure}
    \begin{subfigure}{0.5\textwidth}
        \includegraphics[width=\textwidth]{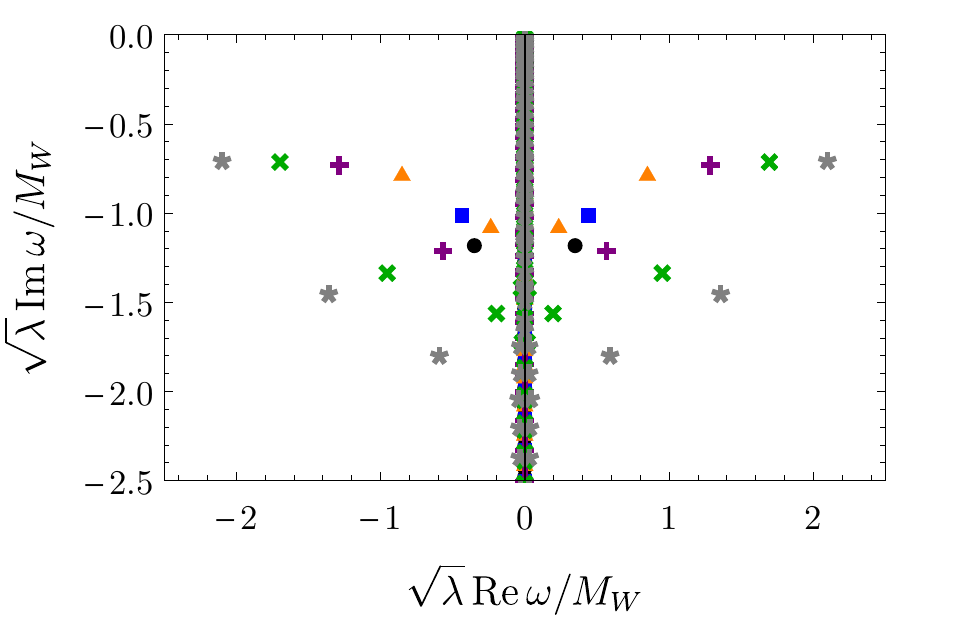}
        \caption{\(\k=10\)}
    \end{subfigure}
    \caption{QNM frequencies, in units of \(M_W/\sqrt{\l}\), for \(l=0\) to \(l=5\) and various values of \(\k\). All results were obtained using Leaver's method with \(300 \times 300\) matrices. For all \(l\), we find many modes on the negative imaginary axis, consistent with a branch cut. We also find pairs of isolated modes with equal and opposite real parts, the number of which increases as we increase \(l\).}
    \label{fig:modes_kappa}
\end{figure}

Moving now to non-zero \(l\), we find that the QNM frequencies gain non-trivial dependence on \(\k\). Fig~\ref{fig:modes_kappa} shows our numerical results for the QNM frequencies for \(l \leq 5\), for various values of \(\k\) between \(\k=1/10\) and \(\k=10\), obtained using Leaver's matrix method with \(M=300\). The qualitative form of each plot is similar. For all \(l\) and \(\k\), we find many modes on the negative imaginary axis, which we expect to coalesce into a branch cut as \(M \to \infty\). Additionally, we find pairs of isolated modes with equal and opposite \(\Re \w\) and with negative \(\Im \w\). The number of such pairs increases as we increase \(l\).

\begin{figure}
    \begin{subfigure}{\textwidth}
    \begin{center}
    \includegraphics[scale=0.8]{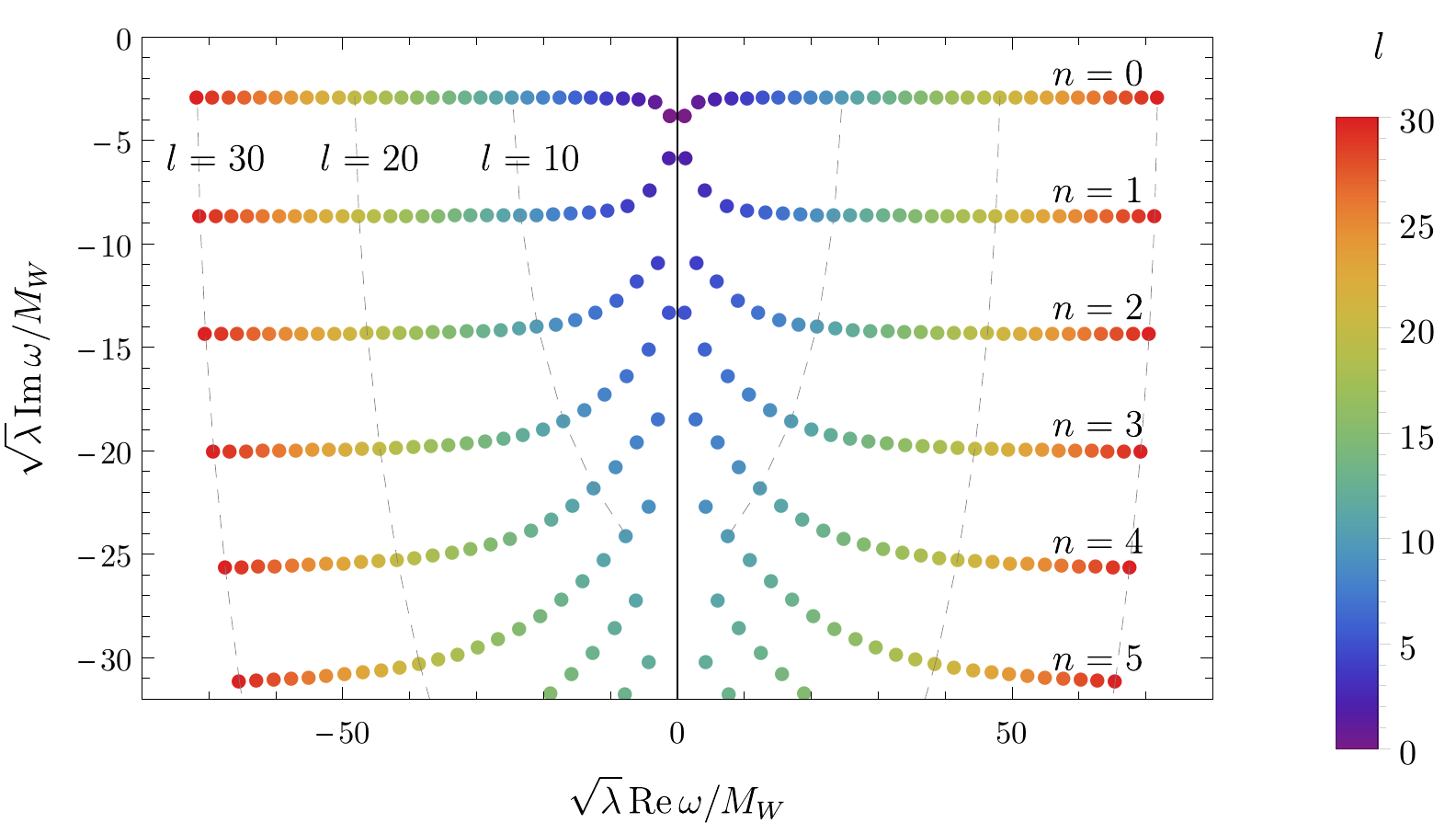}
    \caption{Complex plane, \(\k=1\).}
    \label{fig:kappa1_large_l_complex_plane}
    \end{center}
    \end{subfigure}
    \begin{subfigure}{0.5\textwidth}
        \includegraphics[width=\textwidth]{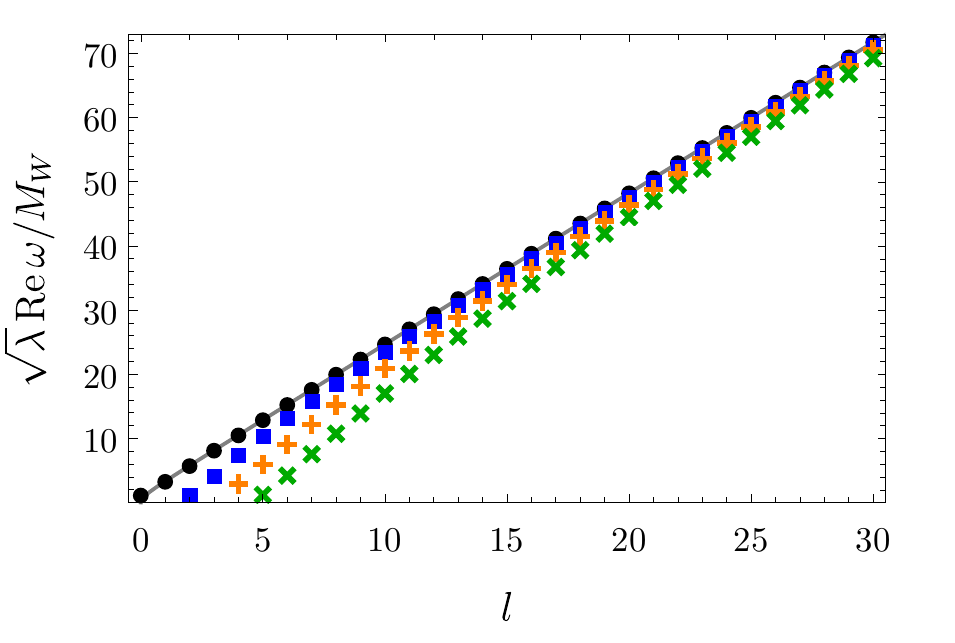}
        \caption{Real part, \(\k=1\).}
        \label{fig:kappa1_large_l_real}
    \end{subfigure}
    \begin{subfigure}{0.5\textwidth}
        \includegraphics[width=\textwidth]{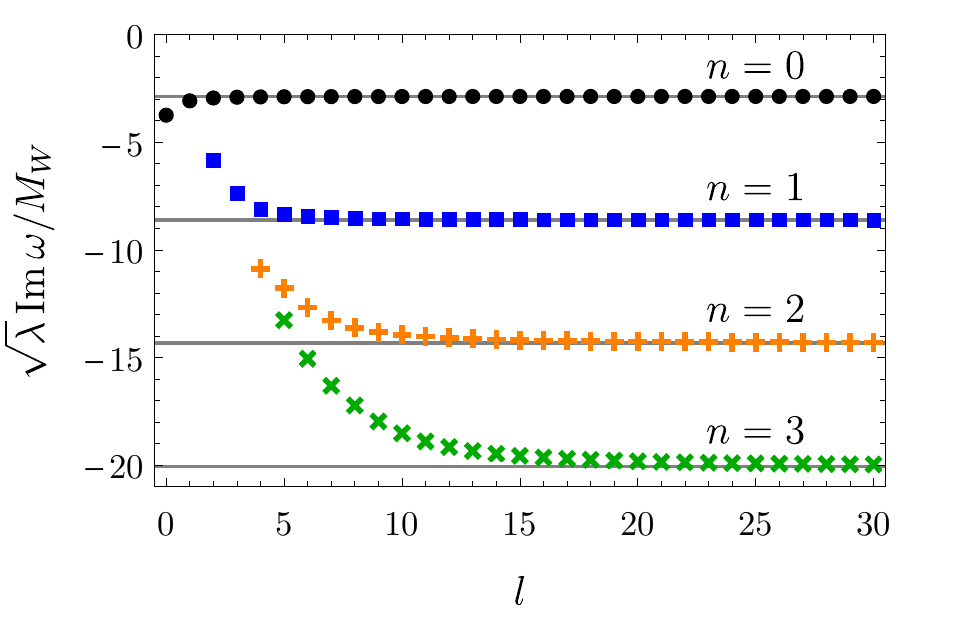}
        \caption{Imaginary part, \(\k=1\).}
        \label{fig:kappa1_large_l_imaginary}
    \end{subfigure}
    \caption{
        \textbf{(a)} QNM frequencies,  in units of \(M_W/\sqrt{\l}\), for \(\k=1\) and \(0 \leq l \leq 30\), all obtained using Leaver's method with \(300 \times 300\) matrices. For clarity, we have not shown the purely imaginary QNM frequencies, which are present for all \(l\). We have labelled the QNM frequencies by an overtone index \(n\). The dashed grey lines are a guide to the eye, connecting QNM frequencies with the same value of \(l\). \textbf{(b,\,c)} The real and imaginary parts of the first four overtones, as functions of \(l\), for \(\k=1\). The grey lines show the WKB approximation in eq.~\eqref{eq:kappa_1_wkb}, which works well for \(l\) sufficiently large compared to \(n\).
    }
    \label{fig:kappa1_large_l}
\end{figure}

The results in fig.~\ref{fig:modes_kappa} show hints of a pattern in the distribution of QNM frequencies at large \(l\). In particular, at fixed \(\k\), for every QNM frequency at some value \(l=j\) with frequency \(\w_{j}\), another QNM frequency appears at \(l=j+1\) with frequency \(\w_{j+1}\), such that \(\Re(\w_{j+1} - \w_j)\) is approximately independent of \(j\), and \(\Im( \w_{j+1} -\w_j)\approx 0\). In other words, the QNM frequencies line up with roughly equal spacing along branches roughly perpendicular to the imaginary axis. This pattern becomes clearer if we plot the QNM frequencies up to a larger value of \(l\). For example, in fig.~\ref{fig:kappa1_large_l_complex_plane} we show the \(\k=1\) QNM frequencies up to \(l=30\). In the figure, we have labelled the QNM frequencies by an overtone index \(n\), where larger \(n\) corresponds to more negative imaginary part.

This pattern can be understood from a first-order WKB approximation\footnote{Although it is important to note that in the present setting, the WKB eikonal approximation and higher order corrections involve nontrivial resolution for $ \omega $ since the QNM frequency and reduced potential do not decouple and the inverse Tortoise transform cannot be found, unlike conventional black hole potentials.}, detailed in appendix~\ref{app:wkb}. In this approximation, valid at large \(l\) and \(|\Re \w \,| \gg |\Im \w \,|\), we find that the QNM frequencies are given by
\begin{equation}
\label{eq:wkb_frequency}
    \w \approx \le[\pm \frac{\sqrt{2} \pi x_0^{5/2}}{1-x_0} \sqrt{l(l+1)} - i\frac{\sqrt{2} \pi x_0^2\sqrt{5 -  3 x_0}}{1-x_0} \le(n + \frac{1}{2}\ri)\ri] \frac{M_W}{\sqrt{\l}},
\end{equation}
where \(x_0\) is the lone root in the range \(x_0 \in[0,1]\) of the quintic equation
\begin{equation} \label{eq:wkb_matching_condition}
    \k^2 x_0^5 - (2-x_0)(1-x_0)^4 = 0.
\end{equation}
When \(l \gg 1\) in eq.~\eqref{eq:wkb_frequency}, the factor of \(\sqrt{l(l+1)}\) in the real part may be approximated simply as \(l\), leading to the equally-spaced real parts described above, while the imaginary part is independent of $l$, but is $\propto- (n+1/2)$. For example, when \(\k=1\), we can numerically solve eq.~\eqref{eq:wkb_matching_condition} to find \(x_0 \approx 0.558921\). Substituting this into eq.~\eqref{eq:wkb_frequency_imaginary}, we obtain
\begin{equation}
\label{eq:kappa_1_wkb}
    \w  \approx \le[ \pm 2.3525 \sqrt{l(l+1)} - 5.7363 \le(n+\frac{1}{2}\ri)i \ri] \frac{M_W}{\sqrt{\l}}.
\end{equation}
In figs.~\ref{fig:kappa1_large_l_real} and~\ref{fig:kappa1_large_l_imaginary} we show the real and imaginary parts of the \(\k=1\) QNM frequencies for \(n \leq 3\) and \(l \leq 30\). The grey lines in the figure show the WKB approximation in eq~\eqref{eq:kappa_1_wkb}, which obviously works very well at large \(l\).

%

\section{Comparisons to Other Systems}
\label{sec:comparisons}

In this section we will compare our results for the QNM spectrum of the $\cN=4$ SYM Coloumb branch soliton in sec.~\ref{sec:results} to three other systems with similar properties: $(3+1)$-dimensional asymptotically flat extremal Reissner-Nordstr\"om black holes, gravastars, and the BPS magnetic monopole of \(SU(2)\) Yang-Mills theory coupled to an adjoint scalar field that breaks the gauge symmetry to \(U(1)\).

\begin{figure}
    \begin{center}
    \includegraphics[scale=0.8]{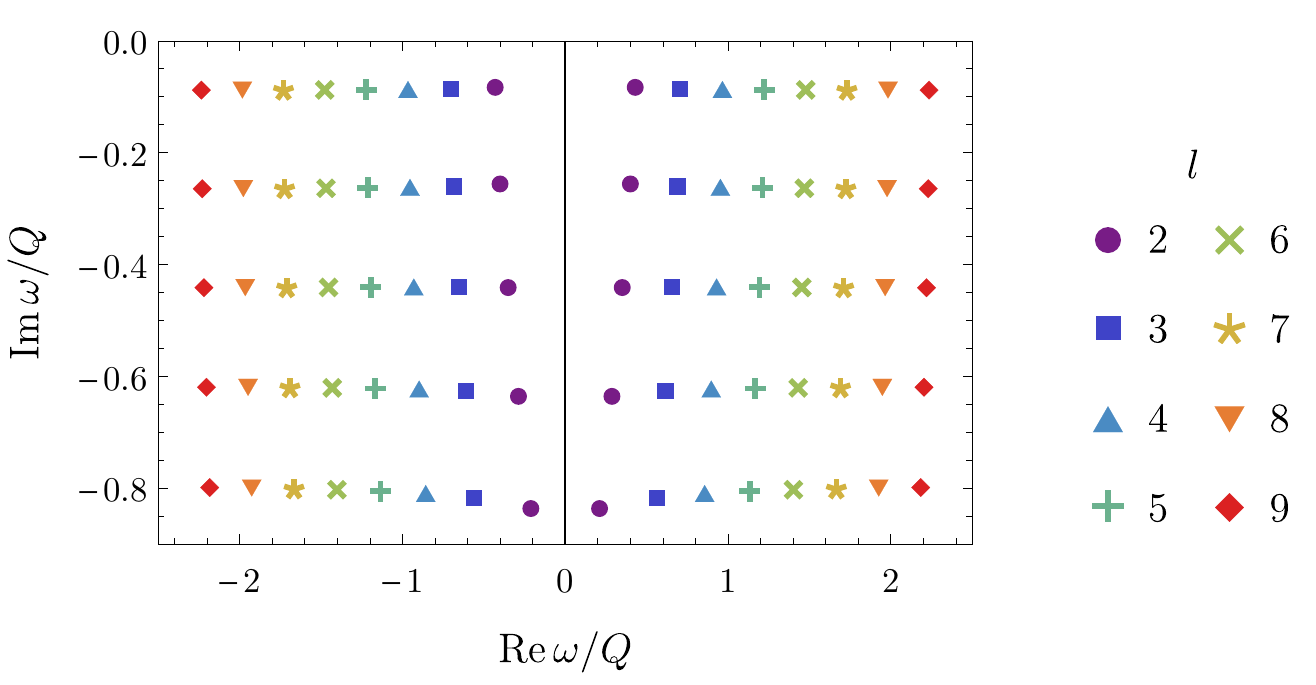}
    \caption{
        QNM frequencies of metric perturbations of the $(3+1)$-dimensional asymptotically flat extremal Reissner-Nordstr\"om black hole, in units of the black hole charge \(Q\), computed using the third-order WKB approximation of refs.~\cite{PhysRevD.35.3621,PhysRevD.35.3632,PhysRevD.37.3378}.
    }
    \label{fig:rn_modes}
    \end{center}
\end{figure}

Fig.~\ref{fig:rn_modes} shows the QNM spectrum of metric perturbations of the extremal Reissner-Nordstr\"om black hole (which are isospectral with the electromagnetic perturbations) up to \(l=9\), computed using the third-order WKB approximation of refs.~\cite{Leaver:1990zz,Onozawa:1995vu,PhysRevD.35.3621,PhysRevD.35.3632,PhysRevD.37.3378,Andersson:1996xw}. How does this QNM spectrum compare to ours? One significant difference is that extremal Reissner-Nordstr\"om's QNM frequencies are all proportional to the black hole's charge \(Q\), whereas our QNM frequencies have non-trivial dependence on the analogous parameter, \(\k\). Another is that the black hole has many QNMs at small \(l\) and the soliton has only a few. Indeed, comparing figs.~\ref{fig:modes_kappa} and~\ref{fig:kappa1_large_l} to fig.~\ref{fig:rn_modes}, the Reissner-Nordstr\"om spectrum is most similar to the low-$\kappa$ spectra in fig.~\ref{fig:modes_kappa} or the large-$l$ spectrum in fig.~\ref{fig:kappa1_large_l}.  In particular, in these cases the QNM frequencies line up with roughly equal spacing along branches roughly perpendicular to the imaginary axis. This similarity is not just qualitative. At large \(l\) and fixed \(n\), the WKB approximation for the black hole QNM frequencies is
\begin{equation}
    \w \approx \le[ \pm \frac{l}{4} - \frac{i}{4 \sqrt{2}} \le(n + \frac{1}{2} \ri)\ri] Q,
\end{equation}
so just as we found for the soliton in the analogous limit, eq.~\eqref{eq:wkb_frequency}, at large $l$ the real part is $\propto \pm l$ while the imaginary part is independent of \(l\), and is $\propto -(n+1/2)$. In fact, such behavior is not unique to Reissner-Nordstr\"om, but occurs generically for asymptotically flat black holes~\cite{Konoplya:2011qq}.

However, one feature of the soliton QNMs is similar to a unique feature of extremal Reissner-Nordstr\"om, namely the late-time behaviour. For a generic asymptotically flat black hole, the Green's functions of massless perturbations exhibit a branch point at the origin of the complex frequency plane, with the branch cut typically oriented along the negative imaginary axis. An important physical consequence of the branch cut is, after a Fourier transform, power law decay of the perturbations at late time $t$~\cite{PhysRevD.5.2419}. For example, massless scalar perturbations of a $(3+1)$-dimensional Schwarzschild or non-extremal Reissner-Nordstr\"om black hole decay as \(t^{-(2l+3)}\), provided the initial data for the perturbation has compact support~\cite{PhysRevD.5.2419,1972GReGr...3..331B,Gundlach:1993tp}. Perturbations of an extremal Reissner-Nordstr\"om black hole, with compact initial data, also decay as \(t^{-(2l+3)}\), \textit{unless} the initial data extends to the horizon, in which case the perturbation decays more slowly, as \(t^{-(2l+2)}\)~\cite{Blaksley:2007ak,Ori:2013iua,Sela:2015vua,Bhattacharjee:2018pqb}.

In each case, the power of \(t\) is determined by the branch cut's strength, which in turn is fixed by the asymptotic behaviour of the Schr\"odinger potential \(V(x_*)\) far from the black hole, at \(x_* \to \infty\)~\cite{Ching:1995tj}. Schwarzschild and Reissner-Nordstr\"om exhibit the same \(t^{-(2l+3)}\) decay because their Schr\"odinger potentials have the same asymptotics at \(x_* \to \infty\). The special feature of extremal Reissner-Nordstr\"om that leads to the \(t^{-(2l+2)}\) decay for perturbations originating at the horizon is that its Schr\"odinger potential vanishes as a power of \(x_*\) near the horizon, \(V(x_*) \sim x_*^{-2}\) as \(x_* \to -\infty\), rather than exponentially, as in Schwarzschild and non-extremal Reissner-Nordstr\"om. As written in eq.~\eqref{eq:schrodinger_Z_asymptotics}, our Schr\"odinger potential also vanishes as a power of \(x_*\) at both \(x_* \to \pm \infty\), albeit as $x_*^{-4}$ rather than Reissner-Nordstr\"om's $x_*^{-2}$. Presumably such power laws arise from the $AdS_2$ throats that appear in both cases. We thus expect perturbations of the soliton to exhibit power-law late-time tails, with a power that depends on the initial data in a fashion similar to extremal Reissner-Nordstr\"om. We leave a detailed analysis of this to future research.

As discussed in sec.~\ref{sec:intro}, in some ways the solitons we consider more closely resemble gravastars than black holes. Unfortunately, we are not aware of any results for QNMs of charged gravastars, so the best comparison we can make is to uncharged ones. The precise QNM spectrum of an uncharged gravastar depends on details such as the thickness of the shell separating the de Sitter bubble inside from the Schwarzschild spacetime outside. However, if a gravastar has either a sufficiently thick shell or is sufficiently compact, its QNMs typically have \(|\Im \w\,|\) much smaller than those of a Schwarzschild black hole of the same mass, with \(|\Im \w\,| \ll |\Re \w\,|\) even at small \(l\)~\cite{Chirenti:2007mk,Pani:2009ss}.\footnote{\(|\Im \w\,| \ll |\Re \w\,|\) also defines a valid regime for comparison with the WKB eikonal approximation.} As fig.~\ref{fig:modes_kappa} shows, at small \(l\) the QNMs of the soliton have \(|\Im \w\,|\) of the same order as \( |\Re \w\,|\) or larger, so in this sense they do not resemble known results for gravastar QNMs.

As also discussed in sec.~\ref{sec:intro}, the $\cN=4$ SYM Coulomb branch soliton is in some ways similar to ``magnetic bags'' formed from a cluster of monopoles in \(SU(2)\) Yang-Mills coupled to an adjoint scalar field that breaks \(SU(2) \to U(1)\)~\cite{Bolognesi:2005rk,Lee:2008ze}, that satisfy $ SU(2) $ BPS system in the Bogomolny limit \cite{Taubes_2014}. To our knowledge no calculation of magnetic bag QNMs has been performed. The best comparison we can make is to a single BPS magnetic monopole, for which the \(l=0\) QNMs of the W-boson field were computed in ref.~\cite{Forgacs:2003yh}. We computed QNMs of only massless fields, whereas the W-bosons are of course massive, so we should not necessarily expect the two spectra to be similar. Indeed, ref.~\cite{Forgacs:2003yh} found no evidence for a branch cut, and moreover found an infinite number of QNM frequencies with non-zero real part, in contrast to the single pair we found for \(l=0\). However, the single monopole does have some similarity to black holes. The QNMs of the \(SU(2)\) monopole may be labelled by an overtone index \(n\), where for \(n \gg 1\) the frequency of the \(n\)-th mode is given by
\begin{equation}
\label{eq:su2_monopole_modes}
    \w_n \approx \le( \pm \sqrt{1 - \frac{1}{n^2}} - \frac{0.2 i}{n^3} \ri) M_W.
\end{equation}
An infinite number of modes thus accumulate near the real frequency axis, with \(|\Re \w\,|\approx M_W\). Eq.~\eqref{eq:su2_monopole_modes} shows that the lifetime of the \(n\)-th mode, \(1/|\Im \w_n\,|\), grows as \(n^3\) for large \(n\). These long-lived, large \(n\) modes cause the amplitudes of spherically symmetric excitations of the \(SU(2)\) monopole to decay as a power law \(t^{-5/6}\) at late time \(t\)~\cite{Forgacs:2003yh,Fodor:2003yg}. Remarkably, the amplitudes of massive perturbations of $(3+1)$-dimensional black holes exhibit the same \(t^{-5/6}\) power law decay, independent of the value of \(l\), the spin of the fluctuating field, or the type of black hole~\cite{Koyama:2000hj,Koyama:2001ee,Moderski:2001tk,Finster:2001vn,Koyama:2001qw,Jing:2004zb,Moderski:2005hf,Konoplya:2006gq}. In the black hole case, the \(t^{-5/6}\) power originates from a branch cut between \(\w = \pm m\), where \(m\) is the mass of the fluctuating field~\cite{Koyama:2001qw}.

\section{Summary and Outlook}
\label{sec:outlook}

We holographically computed the QNM spectrum of a BPS soliton on the Coloumb branch of $\cN=4$ SYM at large $N$ and large coupling, using the dual description in terms of a probe D3-brane in $AdS_5 \times S^5$. Schwarz proposed that these solitons may reproduce some features of $(3+1)$-dimensional asymptotically flat extremal Reissner-Nordstr\"om black holes~\cite{Schwarz:2014rxa,Schwarz:2014zsa}. Our results provide some evidence for this proposal.

Our main results appear in figs.~\ref{fig:l0_modes}, \ref{fig:modes_kappa}, and~\ref{fig:kappa1_large_l}, and the WKB result for large angular momentum in eq.~\eqref{eq:wkb_frequency}. The latter in particular leads to QNM frequencies equally spaced along branches perpendicular to the imaginary axis, with dependence on $l$ and overtone index $n$ of the same form as those of asymptotically flat black holes, as we argued in sec.~\ref{sec:comparisons}. We also argued that because the soliton's fluctuations have an effective Schr\"odinger potential with power-law rather than exponential decay near the soliton, they should exhibit late-time decay with power laws more similar to extremal Reissner-Nordstr\"om than to non-extremal Reissner-Nordstr\"om or Schwarzschild black holes.

Our results suggest several avenues for further research.

Holographic solutions for other BPS solitons are known. For example, solutions are known for probe M5-brane solutions in $AdS_7 \times S^4$ that describe string-like solitons in the $(5+1)$-dimensional $\cN=(2,0)$ supersymmetric CFT~\cite{Schwarz:2014rxa}.\footnote{Such probe M5-brane solutions were found in the full asymptotically flat M5-brane background of 11-dimensional supergravity in ref.~\cite{Gauntlett:1999xz}. The solutions in \(AdS_7 \times S^4\) may be obtained by inverting the sign of a parameter analagous to \(\k\) in the probe M5-brane solutions of ref.~\cite{Rodgers:2018mvq}.} Do these support QNMs, and if so, are they similar to those of other objects, like black strings?

In the magnetically charged case, our solitons are BPS monopoles of the GNO type, and should carry non-Abelian charges under the dual unbroken gauge group. These are not directly visible in the strong coupling picture where the gauge degrees of freedom are replaced with closed strings/gravity. A natural question is whether we can study such monopoles in ${\cal N}=4$ SYM at weak coupling. A potential problem that one may have to contend with whilst analyzing massless fluctuations around  non-Abelian monopoles in ${\cal N}=4$ SYM is that the putative non-Abelian zero modes are not normalizable \cite{Dorey:1995me,Dorey:1996hx}. Furthermore, finite charge multi-monopole solutions at weak coupling are known not to be spherically symmetric whereas our holographic solution describes a spherical dyonic shell.  Therefore the most interesting first step at weak coupling is to simply consider the large charge magnetic bag solutions of Bolognesi~\cite{Bolognesi:2005rk} embedded in ${\cal N}=4$ SYM and compute QNMs at weak coupling. This can first be done in the $SU(2)$ theory broken to $U(1)$ where we do not need to worry about non-Abelian zero modes. We  also remark that the scalar field profile of the bag solution of~\cite{Bolognesi:2005rk}, matches precisely the corresponding profile in our strong coupling description.

We mentioned a number of objects sharing features with extremal Reissner-Nordstr\"om for which QNMs should be computed, including charged gravastars and magnetic bags~\cite{Bolognesi:2005rk}. In a purely QFT calculation (not holographic), Popescu and Shapere found BPS solitons in $\cN=2$ $SU(2)$ pure YM with properties very similar to those of the $\cN=4$ SYM soliton~\cite{Popescu:2001rf}. Does their soliton support QNMs, and if so how does its spectrum compare to ours? Can their solution be generalised to $\cN=4$ SYM, and if so, then is a purely QFT calculation of the QNM spectrum possible? How does it compare to our holographic calculation?

Do any or all of the BPS solitons above have non-zero entropy of any kind, such as entanglement entropy, that scales with their surface area? If so, then what are the microstates, and why do they not scale with volume? Can they teach us anything about the microstates that contribute to a black hole's Bekenstein-Hawking entropy? The probe D3-brane solution we studied is extremely similar to supertube solutions for other probe branes~\cite{Mateos:2001qs,Emparan:2001ux,Hyakutake:2002fk,Hyakutake:2005ka}, for which worldvolume zero modes describing shape deformations produce a non-zero Cardy entropy~\cite{Palmer:2004gu}, which may be related to black hole entropy in string theory~\cite{Lunin:2001jy,Lunin:2002qf,Mathur:2002ie,Mathur:2003hj}. Do the probe D3-brane solutions have a similar entropy from zero modes, and if so, then what kind of entropy is it in $\cN=4$ SYM?

Of course, the over-arching question is: to what extent can horizonless objects, in QFT and gravity, capture the physics of black holes? We intend to pursue this and many of the other questions above in the future, using this paper as a foundation.

\section*{Acknowledgements}

We thank Roman Konoplya, Dmitry Sorokin and Konstantin Zarembo for useful remarks and discussions. We would especially like to thank Adam Chalabi and Jacopo Sisti for collaboration on an early stage of this project. S.~P.~K. acknowledges support from STFC grant ST/P00055X/1. A.~O'B. is a Royal Society University Research Fellow. A. P. is supported by SFI and the Royal Society RGF$\backslash$EA$\backslash$180167. The work of R.~R. was supported by the D-ITP consortium, a program of the Netherlands Organisation for Scientific Research (NWO) that is funded by the Dutch Ministry of Education,  Culture and Science (OCW).

\appendix

\section{Leaver's Method}
\label{app:leavers}

In this appendix we describe Leaver's matrix method~\cite{Leaver:1990zz}, which we use to determine QNM frequencies. Starting from the equation of motion for fluctuations on the $S^5$, eq.~\eqref{eq:S5_EOM}, the first step is to factor the singular behaviour out of \(Z_{lm}(x)\), by defining
\begin{equation}
Z_{lm}(x) \equiv x(1-x) \exp \le(\frac{i \wb}{x} + \frac{i \k \wb}{1-x} \ri) \cZ_{lm}(x).
\end{equation}
The outgoing boundary conditions become the condition that \(\cZ_{lm}(x)\) is regular at \(x=0\) and \(x=1\).
In terms of \(\cZ_{lm}(x)\), eq.~\eqref{eq:S5_EOM} becomes 
\begin{equation}
\label{eq:leavers_eom}
    a(x) \cZ_{lm}''(x) + b(x) \cZ_{lm}'(x) + c(x) \cZ_{lm}(x) = 0,
\end{equation}
with coefficients\footnote{We could of course multiply \(a(x)\), \(b(x)\) and \(c(x)\) by the same arbitrary function of \(x\) without changing the equation of motion. We have used a normalisation that makes all three functions polynomials in \(x\) (of the smallest possible degree), so that their Taylor expansions about \(x=1/2\) contain only a finite number of terms, leading to more convenient forms of the recurrence relations for \(\mathcal{Z}_{lm,I}\) defined in eq.~\eqref{eq:calZ_power_series}.}
\begin{align}
    a(x) &\equiv x^2 (1-x)^2,
    \nonumber \\
    b(x) &\equiv 2 x (1-x)(1-2x) - 2 i \wb \le[1 - 2x - (\k-1) x^2 \ri],
    \nonumber
    \\
    c(x) &\equiv - 2 x(1-x) - l(l+1) x^2 + 2 i \wb \le[1 + (\k-1)x \ri] + 2 \k \wb^2.
\end{align}
We now follow ref.~\cite{Onozawa:1995vu} in writing \(\cZ(x)\) as a power series around \(x=1/2\),
\begin{equation} \label{eq:calZ_power_series}
    \cZ_{lm}(x) = \sum_{I=0}^\infty \le(x - \frac{1}{2} \ri)^I \cZ_{lm,I}.
\end{equation}
Substituting this series into the equation of motion eq.~\eqref{eq:leavers_eom}, we find that the coefficients \(\cZ_{lm,I}\) satisfy a five-term recurrence relation,
\begin{equation} \label{eq:recurrence_relation}
    \a_I \cZ_{lm,I+2} + \b_I \cZ_{lm,I+1} + \g_I \cZ_{lm,I} + \d_I \cZ_{lm,I-1} + \e_I \cZ_{lm,I-2} = 0,
\end{equation}
with coefficients
\begin{align}
    \a_I &= \frac{1}{16}(I+1)(I+2),   \qquad \b_I= \frac{1}{2} i \wb (\k-1)(I+1),
    \nonumber \\[1em]
    \g_I &= - \frac{I^2 + I + 1}{2} - \frac{l(l+1)}{4} + i \wb(\k+1)(2I+1) + 2 \k \wb^2,
    \\[1em]
    \d_{I>0} &= - l(l+1) + 2 i \wb (\k-1)I, \qquad \e_{I>1} = (I+l)(I-l-1),
    \nonumber
\end{align}
and $\delta_0=0$, $\e_0=0$, and $\e_1=0$. We can then write the recurrence relation eq.~\eqref{eq:recurrence_relation} as a matrix equation,
\begin{equation}
\label{eq:leavers_matrix_equation}
    \sum_{J=0}^\infty C_{IJ}(\wb) \cZ_{lm,J} = 0,
\end{equation}
where the infinite-dimensional matrix of coefficients is
\begin{equation}
    C(\wb) = \begin{pmatrix}
        \g_0 & \b_0 & \a_0 & 0 & 0 & 0 & 0 & \dots
        \\
        \d_1 & \g_1 & \b_1 & \a_1 & 0 & 0 & 0 & \dots
        \\
        \e_2 & \d_2 & \g_2 & \b_2 & \a_2 & 0 & 0 & \dots
        \\
        0 & \e_3 & \d_3 & \g_3 & \b_3 & \a_3 & 0 & \dots
        \\
        \vdots  & \vdots & \vdots & \vdots & \vdots & \vdots & \vdots & \ddots
    \end{pmatrix}.
\end{equation}
Non-trivial solutions of eq.~\eqref{eq:leavers_matrix_equation} exist only at the frequencies for which \(\det C(\wb) = 0\). We determine these frequencies numerically by truncating \(C(\wb)\) to its upper left \(M \times M\) block, i.e. restricting to \(I,J \leq M-1\) in eq.~\eqref{eq:leavers_matrix_equation}. The determinant may then be straightforwardly computed, for example in \texttt{Mathematica}, and its zeros determined numerically.

\section{WKB Approximation}
\label{app:wkb}
In this appendix we derive expressions for the QNM frequencies in a first-order WKB approximation. We make some mild assumptions based on what we observe numerically, so we cannot claim to have captured all of the QNMs. However, we do find  good agreement with the numerics. We begin by writing the equation of motion eq.~\eqref{eq:S5_EOM} in the form
\begin{equation}
    Z_{lm}''(x) + \eta(x) Z_{lm}(x) = 0,
    \qquad
    \eta(x) \equiv \frac{\wb^2 \le[ \k^2 x^4 + (1-x)^4 \ri] }{x^4 (1-x)^4} - \frac{L^2}{(1-x)^2},
\end{equation}
where we defined \(L^2 \equiv l(l+1)\) (not to be confused with the \(AdS\) radius \(L\) in eq.~\eqref{eq:background_metric}). The function \(\eta(x)\) has a single turning point in \(x\in[0,1]\), located at \(x = x_0\) determined by

\begin{equation} \label{eq:wkb_turning_point}
    \eta'(x_0) =\frac{
         4 \wb^2 \le[ \k^2 x_0^5 - (1-x_0)^5 \ri] - 2 L^2 x_0^5 (1-x_0)^2
     }{x_0^5 (1-x_0^5)} = 0.
\end{equation}

Standard first-order WKB analysis (see ref.~\cite{Konoplya:2011qq} for a review) leads to the result that the QNM frequencies are those for which
\begin{equation}
\label{eq:wkb_condition}
    n = \frac{- i \eta(x_0)}{\sqrt{2 \eta''(x_0)}} -  \frac{1}{2}
\end{equation}
is a natural number. Directly solving eq.~\eqref{eq:wkb_condition} for \(\wb\) seems impossible. To make progress, we assume, motivated by our numerical results, that at large \(L\) the real and imaginary parts of the frequencies scale as
\begin{equation}
    \wb_R \equiv \Re \wb = \cO(L),
    \quad\quad
    \wb_I \equiv \Im \wb = \cO(L^0).
\end{equation}
If we then expand eq.~\eqref{eq:wkb_condition} for large \(L\) and fixed \(n\), we find
\begin{align}
\label{eq:wkb_condition_expanded}
    n + \frac{1}{2}  &= i\frac{
        (1 - x_0^2) x_0^4 L^2 - f(x_0) \wb_R^2
    }{
        2  x_0 (1-x_0)\sqrt{
            10 \tilde{f}(x_0) \wb_R^2  - 3 (1-x_0)^2 x_0^6 L^2
        }
    }
    \nonumber \\ &\phantom{=}
    +
        \frac{
            x_0^4 (1-x_0)^2 \le[2 \k^2 x_0^6 + (2 x_0^2 -10 x_0 + 5)(1-x_0)^4\ri] L^2
            +5 f(x_0) \tilde{f}(x_0) \wb_R^2
        }{
            x_0 (1-x_0)  \le[  10 \tilde{f}(x_0) \wb_R^2  - 3 (1-x_0)^2 x_0^6 L^2 \ri]^{3/2}
        } \wb_R \wb_I
    \nonumber \\ &\phantom{=}
        + \cO\le(L^{-1}\ri),
\end{align}
where \(f(x) \equiv \k^2 x^4 + (1-x)^4\) and \(\tilde{f}(x) \equiv \k^2 x^6 + (1-x)^6\). The first term on the right-hand side of eq.~\eqref{eq:wkb_condition_expanded} is \(\cO(L)\). Demanding that this term vanishes, we find
\begin{equation} \label{eq:wkb_frequency_real}
    \wb_R = \pm \frac{x_0^2 (1-x_0) L}{\sqrt{\k^2 x_0^4 + (1-x_0)^4}}.
\end{equation}
Substituting this into the second term on the right-hand side of eq.~\eqref{eq:wkb_condition_expanded}, which is \(\cO(L^0)\) and should therefore match the \(n+\frac{1}{2}\) on the left-hand side, we find

\begin{equation}
\label{eq:wkb_frequency_imaginary}
    \wb_I = - i \le(n + \frac{1}{2} \ri) \frac{x_0 (1-x_0)
        \sqrt{
            7 \k^2 x_0^6 + (1-x_0)^4 (7 x_0^2 - 20 x_0 + 10)
        }
    }{
        \k^2 x_0^4 + (1 - x_0)^4
    }.
\end{equation}
If we substitute eq.~\eqref{eq:wkb_frequency_real} for \(\wb_R\) into eq.~\eqref{eq:wkb_turning_point}, we find that at leading order in large \(L\) the turning point satisfies
\begin{equation}
\label{eq:wkb_turning_point_substituted}
    \k^2 x_0^5  - (2-x_0)(1-x_0)^4 = 0.
\end{equation}
We can use this equation to replace the explicit \(\k\)-dependence in our expressions for \(\wb_R\) and \(\wb_I\), obtaining the simpler expressions
\begin{equation}
    \wb_R = \pm\frac{\sqrt{l(l+1)} x_0^{5/2}}{\sqrt{2}(1-x_0)},
    \quad
    \wb_I = - \le(n + \frac{1}{2}\ri) \frac{x_0^2 \sqrt{5 -3 x_0}}{\sqrt{2}(1-x_0)},
\end{equation}
where we have replaced \(L\) with \(\sqrt{l(l+1)}\). Using \(M_W = v\sqrt{\l}/2\pi L\) we then find that the dimensionful frequency \(\w = v \wb/L\) is given by the expression in eq.~\eqref{eq:wkb_frequency}.

For a given value of \(\k\), we must determine \(x_0\) by solving eq.~\eqref{eq:wkb_turning_point_substituted} numerically. However, for some values of \(\k\) we can solve for \(x_0\) exactly, for example \(\k = \sqrt{3}\) corresponds to \(x_0 = 1/2\). We may then obtain approximate solutions by solving for \(x_0\) in an expansion around such a point. One such approximation that works well is to expand around \(\k = \infty\), corresponding to \(x_0 = 0\). Using a Pad\'e approximant in large \(\k\), with numerator and denominator up to \(\cO(\k^0)\), we find
\begin{equation} \label{eq:x0_pade_approximant}
    x_0 \approx  \frac{1 + 6(\k/\sqrt{2})^{2/5}}{1 + (11/2) (\k/\sqrt{2})^{2/5}+5 (\k/\sqrt{2})^{4/5}}.
\end{equation}
Strictly speaking, this approximation should only be valid for large \(\k\), but comparison to the numerical solution for \(x_0\) reveals a less than 1\% error for all \(\k \geq 0\) (although the error in \(\w\) obtained by substituting this result into eq.~\eqref{eq:wkb_frequency} may be larger). For example, setting \(\k = 1\) in eq.~\eqref{eq:x0_pade_approximant} yields \(x_0 \approx 0.55890\), leading to the QNM frequencies \(\w  = \bigl[ \pm 2.3521 \sqrt{l(l+1)} - 5.7356 \bigl(n+\frac{1}{2}\bigr)i \bigr] M_W/\sqrt{\l}\) from eq.~\eqref{eq:wkb_frequency}. This compares well with the numerical solution \(x_0 \approx 0.55892\), with corresponding frequencies \(\w  = \bigl[ \pm 2.3525 \sqrt{l(l+1)} - 5.7363 \bigl(n+\frac{1}{2}\bigr)i \bigr] M_W/\sqrt{\l}\).

\bibliographystyle{JHEP}
\bibliography{soliton_ref}

\end{document}